\documentclass[12pt]{article}
\pdfoutput=1
\usepackage{amsmath}
\usepackage{graphicx}
\usepackage{enumerate}
\usepackage{natbib}
\usepackage{url} 

\newcommand{\blind}{0}

\usepackage{float}
\usepackage{amsmath, amssymb, amsthm}
\usepackage{graphicx}
\usepackage{enumitem}
\usepackage{appendix}
\usepackage{titlesec}
\usepackage[ruled, linesnumbered]{algorithm2e}
\RestyleAlgo{ruled}
\usepackage{rotating}
\usepackage{comment}
\usepackage{caption}
\usepackage{subcaption}

\theoremstyle{definition}

\begin{document}

\def\spacingset#1{\renewcommand{\baselinestretch}%
{#1}\small\normalsize} \spacingset{1}


\if0\blind
{
  \title{\bf Principal Subsimplex Analysis}
  \author{\small Hyeon Lee\thanks{
    This work was partially supported by NSF under Grant DMS-2113404.}\hspace{.2cm}\\
    \small Department of Statistics and Operations Research, University of North Carolina at Chapel Hill\\
    \small Kassel Liam Hingee\thanks{This work was supported by 
    Australian Research Council Discovery Project under  DP220102232.}\hspace{.2cm}\\
    \small Research School of Finance, Actuarial Studies and Statistics, Australian National University\\
    \small Janice L. Scealy\thanks{This work was supported by 
    Australian Research Council Discovery Project under  DP220102232.}\hspace{.2cm}\\
    \small Research School of Finance, Actuarial Studies and Statistics, Australian National University\\
    \small Andrew T. A. Wood\thanks{This work was supported by 
    Australian Research Council Discovery Project under  DP220102232.}\hspace{.2cm}\\
    \small Research School of Finance, Actuarial Studies and Statistics, Australian National University\\
    \small Eric Grunsky\\
    \small Department of Earth and Environmental Sciences, University of Waterloo\\
    \small and \\
    \small J. S. Marron\thanks{
    This work was partially supported by NSF under Grant DMS-2113404.}\\
    \small Department of Statistics and Operations Research, University of North Carolina at Chapel Hill}
  \maketitle
} \fi

\if1\blind
{
  \bigskip
  \bigskip
  \bigskip
  \begin{center}
    {\LARGE\bf Principal Subsimplex Analysis}
\end{center}
  \medskip
} \fi

\bigskip
\begin{abstract}
Compositional data, also referred to as simplicial data, naturally arise in many scientific domains such as geochemistry, microbiology, and economics. In such domains, obtaining sensible lower-dimensional representations and modes of variation plays an important role. A typical approach to the problem is applying a log-ratio transformation followed by principal component analysis (PCA). However, this approach has several well-known weaknesses: it amplifies variation in minor variables; it can obscure important variation within major elements; it is not directly applicable to data sets containing zeros and zero imputation methods give highly variable results; it has limited ability to capture linear patterns present in compositional data. In this paper, we propose novel methods that produce nested sequences of simplices of decreasing dimensions analogous to backwards principal component analysis. These nested sequences offer both interpretable lower dimensional representations and linear modes of variation. In addition, our methods are applicable to data sets contain zeros without any modification. We demonstrate our methods on simulated data and on relative abundances of diatom species during the late Pliocene. Supplementary materials and R implementations for this article are available online.
\end{abstract}

\noindent%
\vfill

\newpage
\spacingset{1.75} 

\section{Introduction}

\subsection{Motivation}
Compositional data, which are also referred to as simplicial data, are multivariate observations consisting of vectors of proportions. Such data are prevalent in domains where the relative magnitude between variables is the primary concern, including fields such as geochemistry (proportions of constituent elements), microbiology (proportions of species), and economics (proportions of portfolio components). Compositional data vectors are characterized by the constraints that the entries are nonnegative and sum to one. This constraint complicates the application of traditional multivariate analysis techniques, which are designed for Euclidean data, to compositional datasets. 

A major challenge with compositional data is identifying meaningful lower dimensional approximations and the corresponding \textit{modes of variation} (see Section 3.1.4. of \citet{marron2021object}.) A mode of variation is a set of data objects (e.g. composition vectors) that is one-dimensional in some sense which describes a component of variation. A prototypical example is found in classical Principal Components Analysis (PCA), where modes of variations are the straight lines through the mean spanned by the loading vectors.

\begin{figure}[t]
    \centering
    \includegraphics[trim = {0 1cm 0 0}, width = 0.3\textwidth]{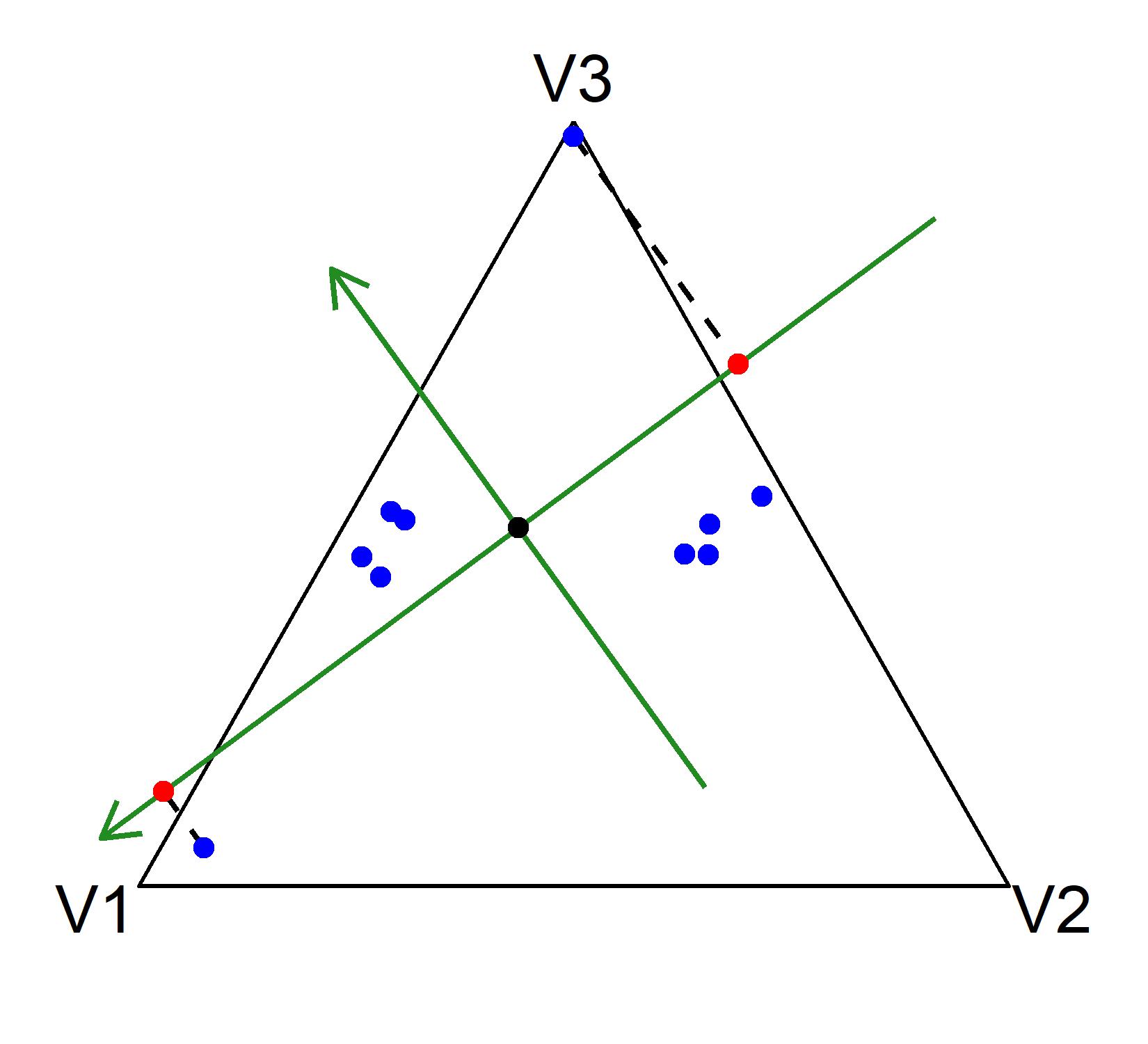}
    \caption{Application of PCA to a 2-dimensional compositional data. Data points (blue dots) and their first two PC directions (green arrows) are represented. Two one-dimensional approximations which fall outside the simplex are indicated by red dots.}
    \label{fig:PCA failure}
\end{figure}

While classical PCA is popular, it often fails to provide effective approximations and modes of variation for non-Euclidean data, including compositional data.
Figure \ref{fig:PCA failure} demonstrates this issue using a compositional dataset with three variables. The distribution of this 2-dimensional compositional data set is effectively visualized by a \textit{ternary plot}, a rotation of the 2-simplex, as shown in Figure \ref{fig:PCA failure}. The blue dots represent data points, while the green arrows depict the first and second principal component directions. The red dots show the one-dimensional approximations of the two selected blue data points, as indicated by the dashed lines. Notably, these red dots fall outside the original compositional space, meaning that the lower-dimensional representations have negative proportions and no longer represents compositions. This example motivates development of more interpretable and effective methods for lower-dimensional approximation in compositional data analysis.

A popular strategy to address such challenges is applying transformations such as log-ratio transformations (\citealt{aitchison1982statistical}, \citeyear{aitchison1983principal}, \citeyear{aitchison1986statistical}) and power transformations (\citealt{aitchison1986statistical}) followed by the usual PCA. Log-ratio transformations bijectively map the interior of the simplex to a Euclidean space of the same dimension, thus enabling application of techniques for Euclidean data. Log-ratio transformations well capture curvature present in compositional data and are the basis of the well-known logistic normal distribution (\citealt{aitchison1986statistical}). On the other hand, power transforms provide a natural continuum between no transformation and log-ratio transformations. Power transforms are effective in normalizing marginal distributions of data, especially remedying high skewness which is common in compositional data.
However, these transformations can strongly distort the original compositional space and lead to undesirable results in some applications. Log-ratio transformations magnify variation within variables with low average proportions (\textit{minor variables}), obscuring important signals present in variables with high average proportions (\textit{major variables}). In addition, log-ratio transformations do not naturally accommodate data points with zero proportions, requiring either a separate treatment of those data points or imputation for zeros. 
\cite{scealy2015robust} proposed a different set of power transformations that map compositional data vectors onto the surface of various manifolds. They then applied classical and robust PCA on the tangent space to the manifold.  
All power transformations inherit the problem of direct PCA in the spirit of Figure \ref{fig:PCA failure}, that the lower dimensional approximations may leave the object space, and the corresponding modes of variation are not interpretable in a compositional sense. 

The above discussion motivates the development of an alternative PCA method that aims to describe variation in major variables while avoiding the limitations of  transformations. In other contexts there has also been a recent growing shift away from transformation based approaches, with a focus on analyzing compositional data  on its original scale of measurement (\citealt{xiong2015virus}, \citealt{Weistuch2022entropy}, \citealt{Fiksel2022regression}, \citealt{scealy2023score}, \citealt{firth2023GEE}, \citealt{Lundborg2023pert}, \citealt{scealy2024robscore}). This article aligns with this view.

\subsection{Backwards Principal Component Analysis}

This paper proposes applying the \textit{backwards Principal Component Analysis} approach to compositional data. Backwards PCA was motivated by \citet{jung2012analysis} then discussed in detail by \citet{damon2014backwards} and \citet[Section 8.6]{marron2021object}. Given an object space that is a rank $d$ manifold, backwards PCA sequentially finds the best-fitting rank $r$ submanifold of the previous rank $r+1$ submanifold. A major advantage of the backwards PCA is that its approximations of any rank naturally remain in the object space, and the nested rank $r$ approximations have natural relationships with the rank $r+1$ approximations which lead to easily defined $r$th scores.
The nested nature of backwards PCA naturally leads to modes of variation, which distinguishes backwards PCA from general dimensionality reduction methods.

Backwards PCA is a unifying framework and requires special attention for each class of object spaces. Backwards PCA has been implemented for a variety of object spaces, including spherical data (\citealt{jung2012analysis}), skeletal representations (\citealt{pizer2013nested}), polyspheres (\citealt{eltzner2015dimension}), nonnegative data (\citealt{zhang2015nested}), and high-dimensional tori (\citealt{eltzner2018torus}, \citealt{zoubouloglou2023scaled}). More discussion on the applications of backwards PCA can be found in Section S1 of the supplementary material.

\subsection{Proposed Methods}

Using the backwards PCA framework, this paper proposes \textit{Principal Subsimplex Analysis (PSA)}. Given a $d$ dimensional compositional data set, PSA effectively identifies a nested sequence of simplices of dimension $r$ for $r=d-1,\cdots,1,0$. The vertices of each subsimplex form a partition of parts, thus PSA is closely related to \textit{amalgamation} in the sense defined in \cite{aitchison1986statistical}. We introduce two versions: \textit{PSA via Simplices} (PSA-S) and \textit{PSA via Orthants} (PSA-O), which use the Euclidean metric and the spherical metric (arc length on the unit sphere), respectively. It will become clear that PSA-S approximates data through amalgamation, while PSA-O combines parts in a related but distinct manner.

Our proposed method has the following four key features.
\begin{enumerate}
    \item \textbf{Targets Variation among Major Variables}: Because PSA uses Euclidean or spherical metrics, it has better ability than transformation-based approaches to capture important variation that occurs among major variables. In addition, PSA is more robust to the existence of pure noise variables.
    \item \textbf{Natural Treatment of Zeros}: By construction of the method, PSA naturally handles data sets with zeros.
    \item \textbf{Compositional Lower Dimensional Approximations}: Each lower dimensional approximation is a simplex whose vertices represent groups of variables, which offers interpretable structure. This approach preserves the important nature of compositional data, such as nonnegativity and potential negative correlations between variables. In particular, the 2-dimensional representation provides useful visualization.
    \item \textbf{Linear Modes of Variation}: PSA produces interpretable modes of variation. In addition, these modes of variation are linear, which serve as a more interpretable alternative to the non-linear modes of variation produced by log-ratio PCA.
\end{enumerate}

\subsection{Related Work} 

\citet{quinn2020amalgams} proposed a method Amalgams which uses amalgamation for dimensionality reduction of compositional data. For $d$-dimensional compositional data and a prescribed number $k<d$, Amalgams searches for a $k$-dimensional amalgamation that optimizes a given objective function, for example, either the average inter-sample distance or classification accuracy. The most important difference between our method and Amalgams is that our method, based on backwards PCA, produces modes of variation.  Our method may also be computationally faster as Amalgams employed a slow genetic algorithm for estimation. 

\subsection{Organization}
The rest of the paper is organized as follows. Section \ref{sec:PSA-S} describes the geometry of simplices and proposes PSA-S. Motivated by scaling challenges, Section \ref{sec:PSA-O} proposes PSA-O. These two versions of PSA, together with Euclidean PCA and the two transformation-based PCAs, are applied to simulated data sets in Section \ref{sec:simulation}, illustrating distinctive properties of our new PSA methods. The same methods are compared on relative abundances of diatom species during the late Pliocene in Section \ref{sec:real data}. The analysis for this article was performed using R Statistical Software (\citealp{r_core}). The R package \texttt{PSA} and the authors' code for this article are available at https://github.com/haneone33/Principal-Subsimplex-Analysis.

\section{Principal Subsimplex Analysis via Simplices}\label{sec:PSA-S}
Each version of the PSA is a special case of backwards PCA that produces a nested sequence of simplices 
\begin{equation}\label{eq:nested simp}
S_d\supset S_{d-1}\supset \cdots \supset S_1\supset S_0
\end{equation} 
such that $S_{r}$ is an $r$-dimensional subsimplex of $S_{r+1}$ for each $r=d-1,\cdots,0$, in the sense formally defined in Section \ref{subsec:simplices}. Each subsimplex $S_r$ is estimated within $S_{r+1}$ so that it best approximates the given data and serves as the rank $r$ approximating subset. Both versions of PSA construct the sequence of simplices based on pairwise aggregation of vertices, but the two versions differ in the way they approximate data points using the subsimplices. 

In this section, we focus on the first version of PSA, \textit{Principal Subsimplex Analysis via Simplices (PSA-S)}. Section \ref{subsec:simplices} defines subsimplices and develops notation for sequences of subsimplices. Section \ref{subsec:pairwise aggr} describes how an aggregation of two vertices yields a subsimplex. Section \ref{subsec:scaling function} handles a scaling problem. The procedure of PSA-S is then given in Section 2.4.

\subsection{Notation}\label{subsec:simplices}

In a Euclidean space $\mathbb{R}^d$, a set $\boldsymbol{V}=\{\boldsymbol{v}_1,\cdots,\boldsymbol{v}_{r+1}\}$ of $r+1$ vectors is said to be \textit{affinely independent} if $(\boldsymbol{v}_1-\boldsymbol{v}_{r+1}),\cdots,(\boldsymbol{v}_r-\boldsymbol{v}_{r+1})$ are linearly independent. An \textit{$r$-dimensional simplex} is the convex hull $S_r$ of an affinely independent set of $r+1$ vectors, that is,
\begin{equation}
S_r=S_r(\boldsymbol{v}_1,\cdots, \boldsymbol{v}_{r+1})=\left\{\sum_{i=1}^{r+1} c_i\boldsymbol{v}_i\in\mathbb{R}^{d+1}: \sum_{i=1}^{r+1}c_i=1, c_1,\cdots,c_{r+1}\ge0\right\}.
\end{equation}
In this case, the vectors $\{\boldsymbol{v}_1,\cdots,\boldsymbol{v}_{r+1}\}$ are called the vertices of $S_r$. Given a simplex $S$, a subsimplex of $S$ is a simplex of any dimension that is also a subset of $S$.

The \textit{$d$-dimensional unit simplex}, denoted by $\Delta_d$, is a special case of a simplex in $\mathbb{R}^{d+1}$ whose vertices are the unit vectors $\boldsymbol{e}_1,\cdots,\boldsymbol{e}_{d+1}$, where $\boldsymbol{e}_j$ is a vector of zeros with a unique one in the $j$th coordinate. A vector of compositions of $d+1$ elements naturally belongs to the $d$-dimensional unit simplex. Given a compositional data set in $\Delta_d$, PSA reveals interpretable lower dimensional representations taking the form of subsimplices of $\Delta_d$ of varying dimensions.

\subsection{Pairwise Aggregation of Vertices}\label{subsec:pairwise aggr}

The sequence of nested subsimplices (\ref{eq:nested simp}) has the following explicit representation using vertices,
\begin{equation}\label{eq:nested simp vertex}
\Delta_d \supset
S_{d-1}\left(\boldsymbol{v}_{1}^{(d-1)},\cdots,\boldsymbol{v}_{d}^{(d-1)}\right) \supset
\cdots \supset
S_{1}\left(\boldsymbol{v}_{1}^{(1)},\boldsymbol{v}_{2}^{(1)}\right) \supset
S_0\left(\boldsymbol{v}_1^{(0)}\right)
\end{equation}
where $\left\{\boldsymbol{v}_{1}^{(r-1)},\cdots,\boldsymbol{v}_{r}^{(r-1)}\right\}$ is an affinely independent subset of $S_{r+1}\left(\boldsymbol{v}_{1}^{(r)},\cdots,\boldsymbol{v}_{r+1}^{(r)}\right)$.
For notational consistency, we set $\boldsymbol{v}_j^{(d)}=\boldsymbol{e}_j, j=1,\cdots,d+1$ so that
\[
\Delta_d = S_d\left(\boldsymbol{e}_1,\cdots,\boldsymbol{e}_{d+1}\right)
=S_d\left(\boldsymbol{v}_1^{(d)},\cdots,\boldsymbol{v}_{d+1}^{(d)}\right).
\]

Given an $r$-simplex $S_r=S_r\left(\boldsymbol{v}_1^{(r)},\cdots,\boldsymbol{v}_{r+1}^{(r)}\right)$, an easy and interpretable way to construct $S_{r-1}\left(\boldsymbol{v}_{1}^{(r-1)},\cdots,\boldsymbol{v}_{r}^{(r-1)}\right)$ is to form the vertex set by aggregating two of the vertices of $S_r$ at a ratio $\alpha_r\in[0,1]$, while keeping the other $r-1$ vertices. Without loss of generality, suppose the two merging vertices are $\boldsymbol{v}^{(r)}_1$ and $\boldsymbol{v}^{(r)}_2$. The new vertex $\boldsymbol{v}_1^{(r-1)}$ is given as the convex combination of these two vertices at an appropriate ratio $\alpha_r\in[0,1]$,
\begin{equation}\label{eq:PSA-S subsimplex 1}
    \boldsymbol{v}_1^{(r-1)} = \alpha_r\boldsymbol{v}_1^{(r)} + (1-\alpha_r)\boldsymbol{v}_2^{(r)}.   
\end{equation}
This new vertex together with the existing vertices after the reindexing
\begin{equation}\label{eq:PSA-S subsimplex 2}
    \boldsymbol{v}_j^{(r-1)}=\boldsymbol{v}_{j+1}^{(r)}, \;j=2,\cdots,r,
\end{equation} 
are affinely independent give an $(r-1)$ subsimplex $S_{r-1}\left(\boldsymbol{v}_1^{(r-1)},\cdots,\boldsymbol{v}_{r}^{(r-1)}\right) \subset S_r$.

\subsection{Scaling Functions}\label{subsec:scaling function}

Given an $r$-dimensional simplex $S_r$, the size of an $(r-1)$-dimensional subsimplex varies depending on its vertex set, in terms of its $(r-1)$-dimensional volume. In particular, a $(d-1)$-dimensional subsimplex of the unit $d$-simplex $\Delta_d$ is smaller than the unit $(d-1)$-simplex unless the vertices of the subsimplex are chosen from the vertices of $\Delta_d$. This raises a challenge in terms of variation explained by the Residual Sum of Squared scores (RSS). This scale problem is handled by correctly identifying an $r$-simplex with the unit $r$-simplex.

An $r$-simplex $S_r$ is naturally identified with the unit $r$-simplex by matching the vertices of $S_r$ to those of the unit $r$-simplex. We define the \textit{scaling function} $\phi_{S_r}$ by
\begin{equation}\label{eq:scale function}
\phi_{S_r}:S_r(\boldsymbol{v}_1,\cdots, \boldsymbol{v}_{r+1})\to\Delta_r, \;
\sum_{j=1}^{r+1} \alpha_j\boldsymbol{v}_j \mapsto (\alpha_1,\cdots,\alpha_{r+1})^\top.     
\end{equation}
Given a sample $\boldsymbol{x}=\sum_{j=1}^{r+1} \alpha_j\boldsymbol{v}_j$,
the proportion associated with vertex $\boldsymbol{v}_j$ is given by $\alpha_j$. In this sense, $\phi_{S_r}$ can be interpreted as writing $\boldsymbol{x}$ in a new coordinate system with basis $\boldsymbol{v}_1,\cdots,\boldsymbol{v}_{r+1}$.

\subsection{Principal Subsimplex Analysis via Simplices (PSA-S)} 
\label{subsec:PSA-S}
 
To describe the PSA-S procedure, assume we have a $d$-dimensional compositional data set $\left\{\boldsymbol{x}_1,\cdots,\boldsymbol{x}_n\right\}$ in $\Delta_d$. We denote by $\hat{\boldsymbol{x}}^{(r)}_i$ the rank $r$ approximation of $\boldsymbol{x}_i$ which is a member of $S_r\left(\boldsymbol{v}^{(r)}_1,\cdots,\boldsymbol{v}^{(r)}_{r+1}\right)$, for $r=d-1,\cdots,0$. For notational consistency, we set $\hat{\boldsymbol{x}}^{(d)}_i=\boldsymbol{x}_i, i=1,\cdots,n$.

\begin{figure}
    \centering
    \begin{subfigure}{0.35\textwidth}
        \centering
        \includegraphics[trim = {0 1cm 0 0}, clip, width = \textwidth]{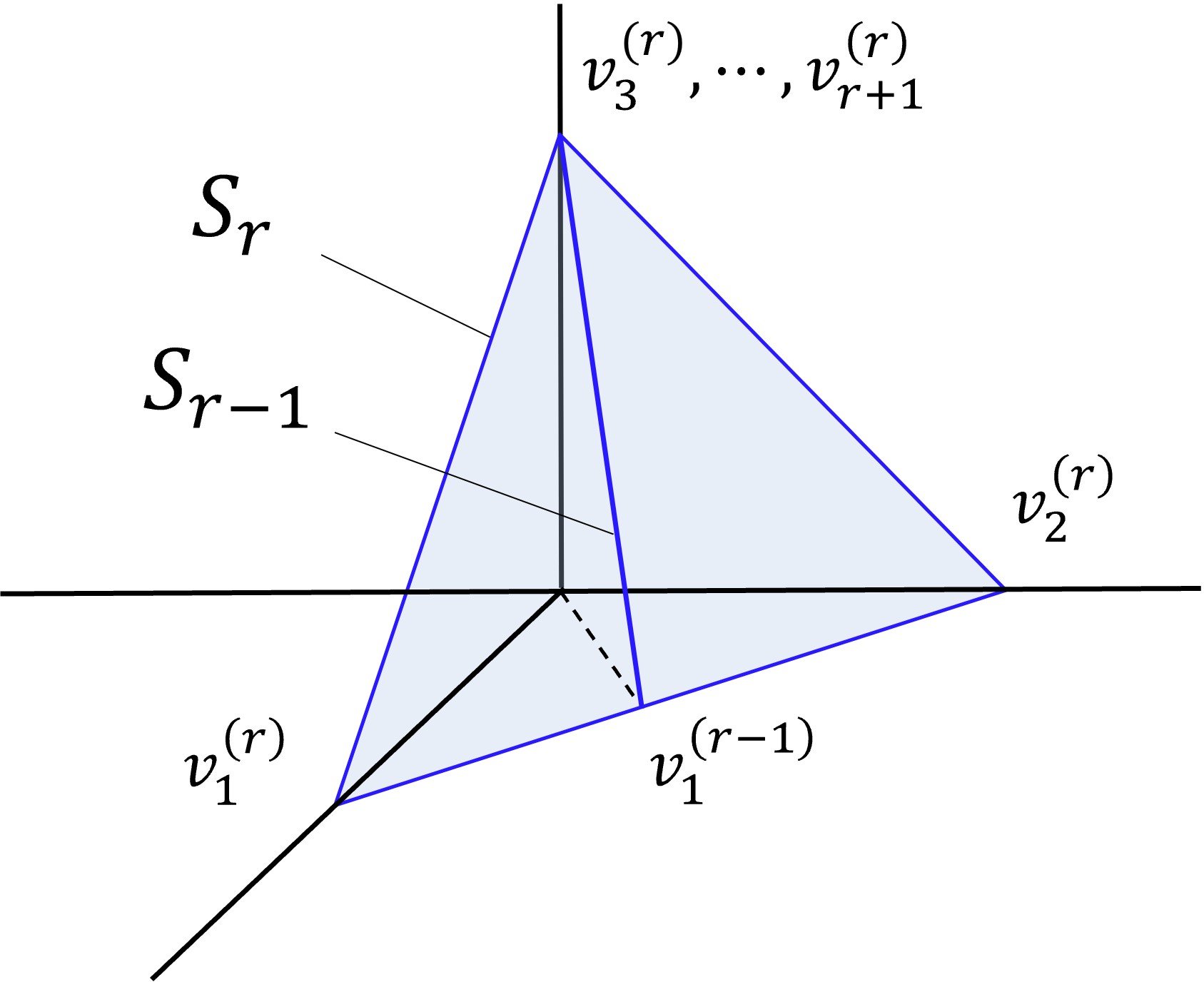}
        \caption{}
    \end{subfigure}
    \begin{subfigure}{0.35\textwidth}
        \centering
        \includegraphics[trim = {0 1cm 0 0}, clip, width = \textwidth]{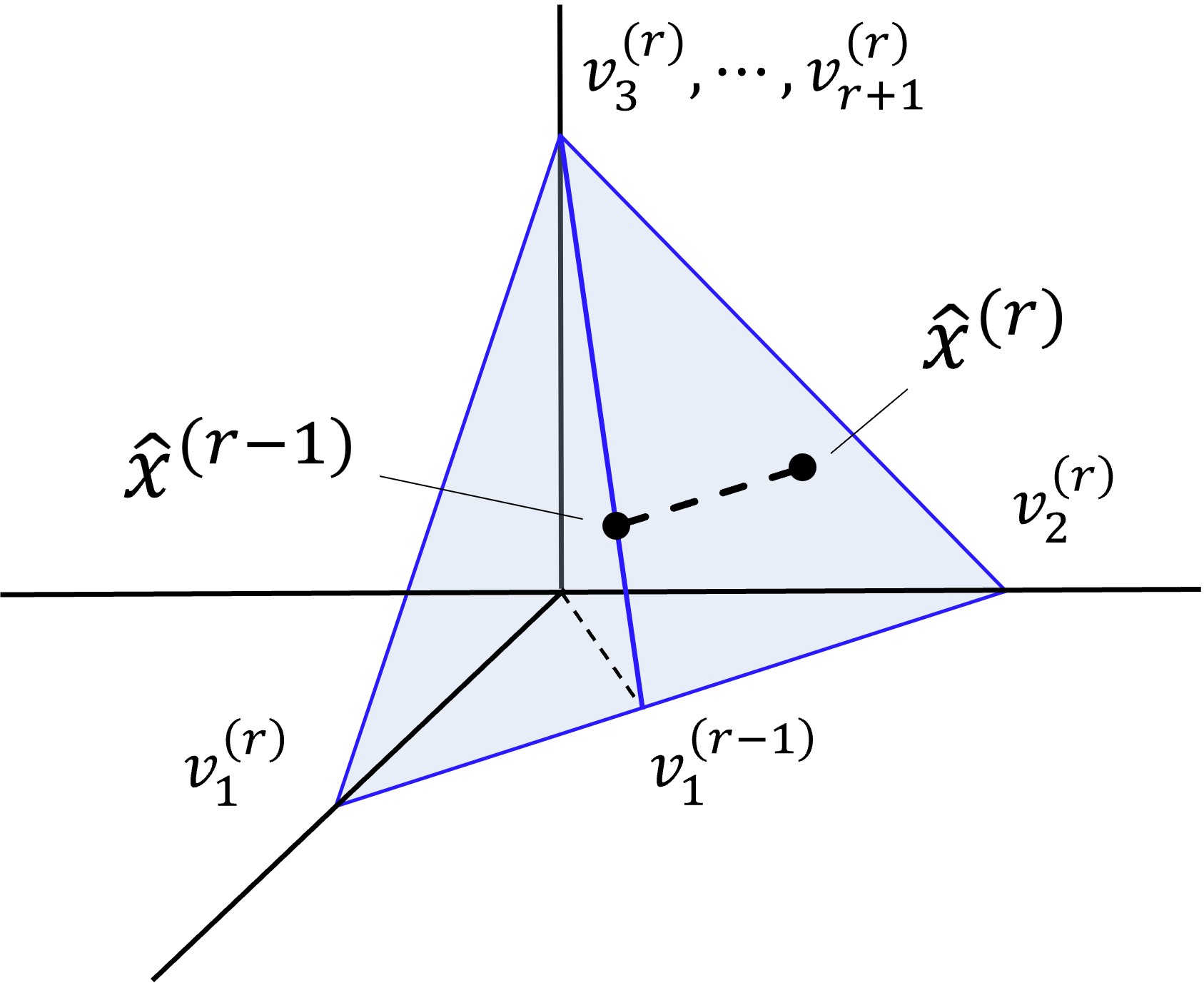}
        \caption{}
    \end{subfigure}
    \caption{An iteration of PSA-S. (a) PSA-S merges two vertices to construct a subsimplex $S_{r-1}$ of $S_r$. (b) PSA-S produces the rank $(r-1)$ approximation $\hat{\boldsymbol{x}}^{(r-1)}$ by transporting $\hat{\boldsymbol{x}}^{(r)}$ parallel to $\boldsymbol{v}^{(r)}_2 - \boldsymbol{v}^{(r)}_1$.}
    \label{fig:PSA-S}
\end{figure}

The subsimplices in (\ref{eq:nested simp vertex}) are constructed inductively through repeated aggregation of vertices. Suppose we have obtained the rank $r$ subsimplex $S_{r}=S_r\left(\boldsymbol{v}^{(r)}_1,\cdots,\boldsymbol{v}^{(r)}_{r+1}\right)$ and the approximations $\hat{\boldsymbol{x}}^{(r)}_i$. Further suppose that the approximating subsimplex $S_{r-1}=S_{r-1}\left(\boldsymbol{v}^{(r-1)}_1,\cdots,\boldsymbol{v}^{(r-1)}_{r}\right)$ is formed by combining $\boldsymbol{v}^{(r)}_1$ and $\boldsymbol{v}^{(r)}_2$ at a ratio $\alpha_r$. (The first two vertices are chosen for notational convenience.) PSA-S approximates a data point $\hat{\boldsymbol{x}}^{(r)}=\sum_{j=1}^{r+1}\hat{x}^{(r)}_j\boldsymbol{v}^{(r)}_j$ in $S_r$ in a \textit{mass preserving} way, as illustrated in Figure \ref{fig:PSA-S}. The point $\hat{\boldsymbol{x}}^{(r)}$ in $S_r$ is mapped onto $S_{r-1}$ along the line segment parallel to the edge connecting $\boldsymbol{v}_1^{(r)}$ and $\boldsymbol{v}_2^{(r)}$. This mapping preserves the weights $x_3,\cdots,x_{r+1}$ and only re-distributes the first two weights according to the ratio $\alpha_r$. In other words, the rank $r-1$ approximation is given by the map $\pi^{PSA-S}_r : S_r\left(\boldsymbol{v}_1^{(r)},\cdots,\boldsymbol{v}_{r+1}^{(r)}\right)\to S_{r-1}\left(\boldsymbol{v}_1^{(r-1)},\cdots,\boldsymbol{v}_{r}^{(r-1)}\right)$,
\begin{equation}
\pi^{PSA-S}_r : \sum_{j=1}^{r+1}\hat{x}^{(r)}_j\boldsymbol{v}^{(r)}_j \mapsto \alpha_r(\hat{x}^{(r)}_1+\hat{x}^{(r)}_2)\boldsymbol{v}^{(r)}_1+(1-\alpha_r)(\hat{x}^{(r)}_1+\hat{x}^{(r)}_2)\boldsymbol{v}^{(r)}_2 +\sum_{j=3}^{r+1}\hat{x}^{(r)}_j\boldsymbol{v}^{(r)}_j .
\end{equation}
Thus the rank $(r-1)$ approximation is $\hat{\boldsymbol{x}}^{(r-1)}_i=\pi^{PSA-S}_r\left(\hat{\boldsymbol{x}}^{(r)}_i\right)$. Notably, the mapping is well defined for data points with zero proportions.

The $r$th score of a point $\boldsymbol{x}_i=\sum_{j=1}^{r+1}\hat{x}^{(r)}_{i,j}\boldsymbol{v}^{(r)}_j$ is the signed distance from $\hat{\boldsymbol{x}}^{(r)}_i$ to $\hat{\boldsymbol{x}}^{(r-1)}_i$ where $S_r$ is rescaled to have unit size through $\phi_{S_r}$:
\begin{equation*}
\begin{split}
s^{(r)}_i:=s^{(r)}(\boldsymbol{x}_i)
&:=sgn\left(\hat{x}^{(r)}_{i,1}-\hat{x}^{(r-1)}_{i,1}\right)\cdot\left\Vert\phi_{S_r}\left(\hat{\boldsymbol{x}}^{(r)}_i\right)-\phi_{S_r}\left(\hat{\boldsymbol{x}}^{(r-1)}_i\right)\right\Vert_2\\
&=\sqrt{2}\Big(-(1-\alpha_r)\hat{x}^{(r)}_{i,1}+\alpha_r \hat{x}^{(r)}_{i,2}\Big).
\end{split}
\end{equation*}
To arrive at an analogue of modes of variation in PCA, define the $r$th loading vector $\boldsymbol{l}_r$ to be the difference of the two merged vertices, i.e.  $\boldsymbol{l}_r=\boldsymbol{v}^{(r)}_2-\boldsymbol{v}^{(r)}_1$. This loading vector indicates the direction pointing from each $\hat{\boldsymbol{x}}^{(r)}_i$ to $\hat{\boldsymbol{x}}^{(r-1)}_i$. The score $s^{(r)}_i$ represents the amount of the movement in that direction because
\begin{equation*}
\begin{split}
\hat{\boldsymbol{x}}^{(r)}_{i} -\hat{\boldsymbol{x}}^{(r-1)}_{i} =
\Big((1-\alpha_r)\hat{x}^{(r)}_{i,1}-\alpha_r \hat{x}^{(r)}_{i,2}\Big)\boldsymbol{v}^{(r)}_1
-\Big((1-\alpha_r)\hat{x}^{(r)}_{i,1}-\alpha_r \hat{x}^{(r)}_{i,2}\Big)
\boldsymbol{v}^{(r)}_2 = \frac{s^{(r)}_i}{\sqrt{2}}\boldsymbol{l}_r.
\end{split}
\end{equation*}
As described in Section 3 of the supplementary material, the corresponding mode of variation is $\left\{\boldsymbol{v}^{(0)}+\frac{s^{(r)}_i}{\sqrt{2}}\boldsymbol{l}_r: i = 1, ..., n\right\}$ and thus $\boldsymbol{l}_r$ is indeed the direction of the mode of variation. More detailed investigation of modes of variation for PSA and backwards PCA are found in the same section of the supplementary material.

The merged vertices and the ratio $\alpha_r$ are chosen so that they minimize the Residual Sum of Squared scores (RSS) $\sum_{i=1}^n \left(s^{(r)}_i\right)^2$. Given the vertices $\boldsymbol{v}_1$ and $\boldsymbol{v}_2$ to be merged, the optimal $\alpha_r$ has a closed-form solution
\begin{equation*}
\alpha_r = \frac{\sum_{i=1}^n x_{i,1}(x_{i,1}+x_{i,2})}{\sum_{i=1}^n(x_{i,1}+x_{i,2})^2}.
\end{equation*}
The pair of vertices are chosen so that RSS is minimized at the optimal $\alpha_r$.

The procedure of PSA-S is summarized in Algorithm 1 of the Supplementary material.

\section{Principal Subsimplex Analysis via Orthants}\label{sec:PSA-O}
PSA-S uses scaling functions to handle the varying size of subsimplices. This rescaling can be avoided by representing unit simplices as unit nonnegative orthants, which are closely related to simplices through the invertible projection map which is described below. PSA-O employs a collection of nonnegative orthants, playing a parallel role to the collection of subsimplices in PSA-S. Because two nonnegative orthants of the same dimension have the same size, (that is, they are isometric Riemannian manifolds with boundary,) scales of approximating subsets in the sequence naturally remain the same.

\subsection{Nonnegative Orthants}

Let $\boldsymbol{V}=\left\{\boldsymbol{v}_1,\cdots,\boldsymbol{v}_{r+1}\right\}$ be a set of $r+1$ orthonormal vectors in $\mathbb{R}^d$. The \textit{$r$-dimensional nonnegative orthant} with the vertex set $\boldsymbol{V}$ is the set $O_r$ of linear combinations of $\boldsymbol{v}_1,\cdots,\boldsymbol{v}_{r+1}\in\mathbb{R}^{r}$ with nonnegative coefficients whose squared sum is one. In other words,
\[
O_r=O_r\left(\boldsymbol{v}_1,\cdots,\boldsymbol{v}_{r+1}\right) 
= \left\{\sum_{i=1}^{r+1}c_t\boldsymbol{v}_i\in\mathbb{R}^{r}:\sum_{i=1}^{r+1}c_i^2=1, c_1,\cdots,c_{r+1}\ge0\right\}.
\]
Note that $O_r$ is a subset of the $(d-1)$-dimensional unit sphere. Important subsets of an $r$-nonnegative orthant $O_r$ are \textit{suborthants}, which are subsets of $O_r$ that are $r'$-nonnegative orthants for some $r'<r$. The \text{unit $d$-nonnegative orthant} $\mathcal{O}_d$ is a special case of a nonnegative orthant in $\mathbb{R}^{d+1}$ whose vertices are $\boldsymbol{e}_1,\cdots,\boldsymbol{e}_{d+1}$. 

For each $r$-nonnegative orthant $O_r\left(\boldsymbol{v}_1,\cdots,\boldsymbol{v}_{r+1}\right)$ there exists a corresponding $r$-simplex $S_r\left(\boldsymbol{v}_1,\cdots,\boldsymbol{v}_{r+1}\right)$ with the same vertex set. Each nonnegative orthant and its corresponding simplex are intimately related through the invertible projection map $P$,
\[
P:S_r\left(\boldsymbol{v}_1,\cdots,\boldsymbol{v}_{r+1}\right)\to O_r\left(\boldsymbol{v}_1,\cdots,\boldsymbol{v}_{r+1}\right),\;
\boldsymbol{x}\mapsto\frac{\boldsymbol{x}}{\Vert\boldsymbol{x}\Vert_2}.
\]
The inverse correspondence is
\[
P^{-1}:O_r\left(\boldsymbol{v}_1,\cdots,\boldsymbol{v}_{r+1}\right)\to
S_r\left(\boldsymbol{v}_1,\cdots,\boldsymbol{v}_{r+1}\right),\;
\boldsymbol{x}\mapsto\frac{\boldsymbol{x}}{\Vert\boldsymbol{x}\Vert_1}.
\]
This projection is equivalent to transporting $\boldsymbol{x}\in S_r$ along the line from the origin through $\boldsymbol{x}$ until it intersects with the unit sphere.

The (geodesic) distance along the surface of the sphere $S^{d-1}$ between two points $\boldsymbol{x}, \boldsymbol{y}$ in an $r$-nonnegative orthant $O_r\left(\boldsymbol{v}_1,\cdots,\boldsymbol{v}_{r+1}\right)$ is the length of the shorter arc that connects two points, which is a part of a great circle. The distance function is given by $d(\boldsymbol{x, y})=\cos^{-1}(\boldsymbol{x^\top y})$. Equipped with this geodesic distance, any two $r$-nonnegative orthants are isometric.

\subsection{Principal Suborthant Analysis (PSA-O)} \label{subsec:PSA-O}
PSA-O proceeds in a manner similar to that of PSA-S but the fundamental low-rank approximation happens in the unit nonnegative orthant $\mathcal{O}_d=O_d(\boldsymbol{e}_1,\cdots,\boldsymbol{e}_{d+1})$. PSA-O has three steps: (i) project data points $\boldsymbol{x}_i$ in $\Delta_d(\boldsymbol{e}_1,\cdots,\boldsymbol{e}_{d+1})$ to $\tilde{\boldsymbol{x}}_i$ in $O_d(\boldsymbol{e}_1,\cdots,\boldsymbol{e}_{d+1})$ through the projection $\boldsymbol{x}\mapsto \frac{\boldsymbol{x}}{\Vert\boldsymbol{x}\Vert_2}$, (ii) find a nested sequence of nonnegative orthants $O_r(\tilde{\boldsymbol{v}}^{(r)}_1,\cdots,\tilde{\boldsymbol{v}}^{(r)}_{r+1})$ and lower dimensional approximations $\tilde{\boldsymbol{x}}_r$ for $r=d,d-1,\cdots,0$, and (iii) project $O_r(\tilde{\boldsymbol{v}}^{(r)}_1,\cdots,\tilde{\boldsymbol{v}}^{(r)}_{r+1})$ and $\tilde{\boldsymbol{x}}_r$ in $O_d(\boldsymbol{e}_1,\cdots,\boldsymbol{e}_{d+1})$ back onto $\Delta_d(\boldsymbol{e}_1,\cdots,\boldsymbol{e}_{d+1})$ through the inverse of the projection map, $\tilde{\boldsymbol{x}}\mapsto \frac{\tilde{\boldsymbol{x}}}{\Vert\tilde{\boldsymbol{x}}\Vert_1}$. As mentioned earlier, the use of nonnegative orthants naturally avoids the scaling problem because all nonnegative orthants of the same dimension have the same size. We will use $\tilde{\boldsymbol{x}}$ to indicate points on nonnegative orthants, and continue to use $\boldsymbol{x}$ and $\hat{\boldsymbol{x}}$ for points on simplices.

The above three steps are detailed as follows. Firstly, we project $\boldsymbol{x}_i$ on $\Delta_d$ to the corresponding point $\tilde{\boldsymbol{x}}_i^{(d)}=\frac{\hat{\boldsymbol{x}}_i^{(d)}}{\Vert\hat{\boldsymbol{x}}_i^{(d)}\Vert_2}$ on $O_r$ and initialize  $\tilde{\boldsymbol{v}}_j^{(d)}=\boldsymbol{e}_j$ for $j=1,\cdots,d+1$ so that
\[
O_d\left(\tilde{\boldsymbol{v}}_1^{(d)},\cdots\tilde{\boldsymbol{v}}_{d+1}^{(d)}\right)
=O_d(\boldsymbol{e}_1,\cdots,\boldsymbol{e}_{d+1}).
\]

Next, similar to the PSA-S algorithm, PSA-O iteratively combines two vertices at some ratio $\alpha_r\in[0,1]$. Suppose the rank $r$ approximating orthant $O_r\left(\tilde{\boldsymbol{v}}_1^{(r)},\cdots\tilde{\boldsymbol{v}}_{r+1}^{(r)}\right)$ has been computed and $\tilde{\boldsymbol{v}}_1^{(r)}$ and $\tilde{\boldsymbol{v}}_2^{(r)}$ are merged to give the new vertex
\begin{equation}
\tilde{\boldsymbol{v}}_{1}^{(r-1)}=\frac{\alpha_r\boldsymbol{e}_{1}+\left(1-\alpha_r\right)\boldsymbol{e}_{2}}{\left\Vert \alpha_r\boldsymbol{e}_{1}+\left(1-\alpha_r\right)\boldsymbol{e}_{2}\right\Vert_{2}}
\in O_r\left(\tilde{\boldsymbol{v}}_1^{(r)},\cdots\tilde{\boldsymbol{v}}_{r+1}^{(r)}\right).
\label{eq:CombineVerticesOrthant}
\end{equation}
This new vertex and the remaining vertices $\tilde{\boldsymbol{v}}_{j}^{(r-1)}=\tilde{\boldsymbol{v}}_{j+1}^{(r)}$ for $j=2,\cdots,d$ form an orthonormal set and thus defines an $(r-1)$-nonnegative suborthant 
\[
O_{r-1}\left(\tilde{\boldsymbol{v}}_1^{(r-1)},\cdots,\tilde{\boldsymbol{v}}_r^{(r-1)}\right). 
\] 

\begin{figure}
    \centering
    \begin{subfigure}{0.35\textwidth}
        \centering
        \includegraphics[trim = {0 1cm 0 0}, clip, width = \textwidth]{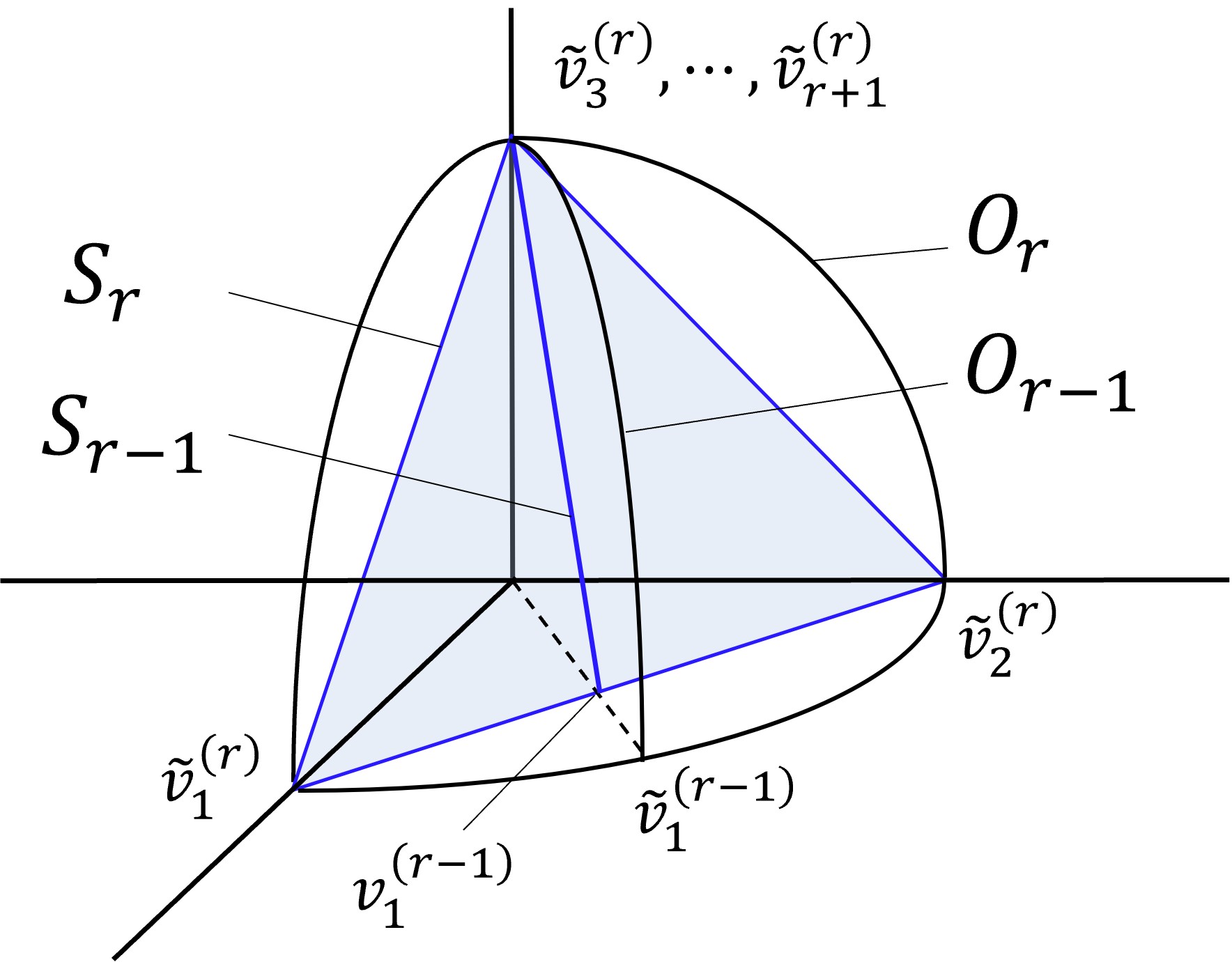}
        \caption{}
    \end{subfigure}
    \begin{subfigure}{0.35\textwidth}
        \centering
        \includegraphics[trim = {0 1cm 0 0}, clip, width = \textwidth]{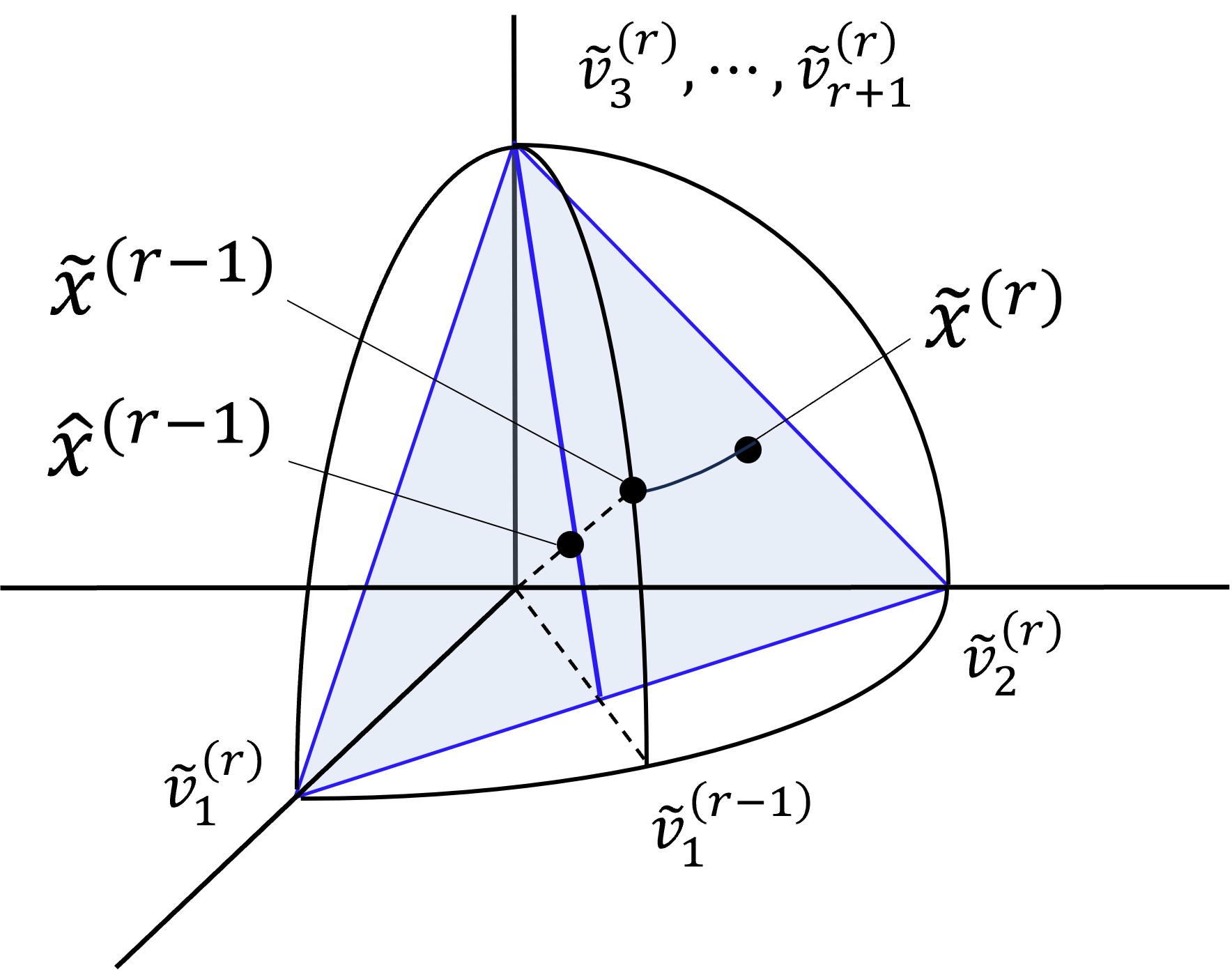}
        \caption{}
    \end{subfigure}
    \caption{An iteration of PSA-O. (a) PSA-O merges two vertices to construct a suborthant $O_{r-1}$ and projects it onto $S_r$ to yield a subsimplex $S_{r-1}$. (b) PSA-O produces $\tilde{\boldsymbol{x}}^{(r-1)}$ by transporting $\tilde{\boldsymbol{x}}^{(r)}$ onto $O_{r-1}$ along the shortest geodesic. Its projection onto $S_{r-1}$ provides $\hat{\boldsymbol{x}}^{(r-1)}$ which lies on the simplex $S_{r-1}$.}
    \label{fig:PSA-O}
\end{figure}

The lower dimensional approximation $\tilde{\boldsymbol{x}}_i^{(r-1)}$ of $\tilde{\boldsymbol{x}}_i^{(r)}\in O_r\left(\tilde{\boldsymbol{v}}_1^{(r)},\cdots\tilde{\boldsymbol{v}}_{r+1}^{(r)}\right)$ is the point in $O_{r-1}\left(\tilde{\boldsymbol{v}}_1^{(r-1)},\cdots,\tilde{\boldsymbol{v}}_r^{(r-1)}\right)$ that is closest to $\tilde{\boldsymbol{x}}_i^{(r)}$ with respect to the geodesic distance (shown as an arc in Figure \ref{fig:PSA-O}(b)) on $\mathbb{S}^d$ and is given by
\[
\tilde{\boldsymbol{x}}^{(r-1)}_i
= \frac{\tilde{\boldsymbol{x}}_i^{(r)}-(\tilde{\boldsymbol{x}}_i^{(r)\top} \boldsymbol{v}_1)\boldsymbol{v}_1}{\sin\left(\cos^{-1}(\tilde{\boldsymbol{x}}_i^{(r)\top} \boldsymbol{v}_1)\right)}.
\]
This projection is equivalent to transporting $\tilde{\boldsymbol{x}}_i^{(r)}$ onto the suborthant $O_{r-1}\left(\tilde{\boldsymbol{v}}_1^{(r-1)},\cdots,\tilde{\boldsymbol{v}}_r^{(r-1)}\right)$ along the great circle that is perpendicular to the suborthant (Figure \ref{fig:PSA-O}). The score is the signed geodesic distance,
\[
s_i^{(r)}:=s^{(r)}(\boldsymbol{x}_i)
=\cos^{-1}\left(\tilde{\boldsymbol{x}}_i^{(r)\top}\tilde{\boldsymbol{x}}_i^{(r-1)}\right).
\]
Again, modes of variation are based on the $r$th loading vector defined as $\boldsymbol{l}_r=\boldsymbol{v}^{(r)}_2-\boldsymbol{v}^{(r)}_1$. Note that although the difference $\hat{\boldsymbol{x}}^{(r)}_i-\hat{\boldsymbol{x}}^{(r-1)}_i$ is not exactly parallel to the loading vector as in the case of PSA-S, the loading vectors still serve as effective representatives of the directions.

Lastly, approximating subsets $O_r(\tilde{\boldsymbol{v}}^{(r)}_1,\cdots,\tilde{\boldsymbol{v}}^{(r)}_{r+1})$ and lower dimensional approximations $\hat{\boldsymbol{x}}^{(r)}_i$ for $r=d,d-1,\cdots,0$ are mapped onto $\Delta_r$ through the inverse of the projection map.

The search for the optimal pair of vertices and the ratio $\alpha_r$ is implemented by multiple grid searches for $\alpha_r$, for each of $r(r+1)/2$ pairs of vertices. Note that the grid search is fast and effective because for each pair of vertices, $\alpha_r$ is a one-dimensional parameter in the bounded region [0,1].

The procedure of PSA-O is summarized in Algorithm 2 of the Supplementary material.

\section{Benchmark Methods}

Both the simulation study in Section \ref{sec:simulation} and the applications to real data in Section \ref{sec:real data} compare two versions of PSA with three common benchmark approaches, Euclidean PCA, power transform PCA, and log-ratio PCA. To describe the transformations, we use interchangeably use $\boldsymbol{x}$ and $(x_i)$ to denote a vector in $\mathbb{R}^D$.

Power transform PCA starts with transformation $(x_i)\mapsto \left(x_i^{\alpha}\right)$ for some parameter $\alpha$. In this paper, $\alpha=1/2$ was used for all data sets for simplicity. For improved analysis, $\alpha$ can be selected to maximize the  Gaussian likelihood of the transformed data.

For log-ratio PCA, the central log-ratio transformation $clr$ was used where
\begin{equation}
clr:(x_i)\mapsto \log\left(\frac{x_i}{g(\boldsymbol{x})}\right) 
\end{equation}
and $g(\boldsymbol{x})=\sqrt[D]{x_1\cdots x_D}$ is the geometric mean of the entries of $\boldsymbol{x}$. For the $clr$ map to be defined, zeros in the data set should be removed or replaced. 
In this paper, only for log-ratio PCA, zeros were replaced by half of the overall minimum nonzero value of the data.

\section{Simulation Studies} \label{sec:simulation}

In this section, the behavior of the two versions of PSA and the three benchmark methods are compared in two toy examples.

\begin{figure}[t]
    \centering
    \includegraphics[width = 0.8\textwidth]{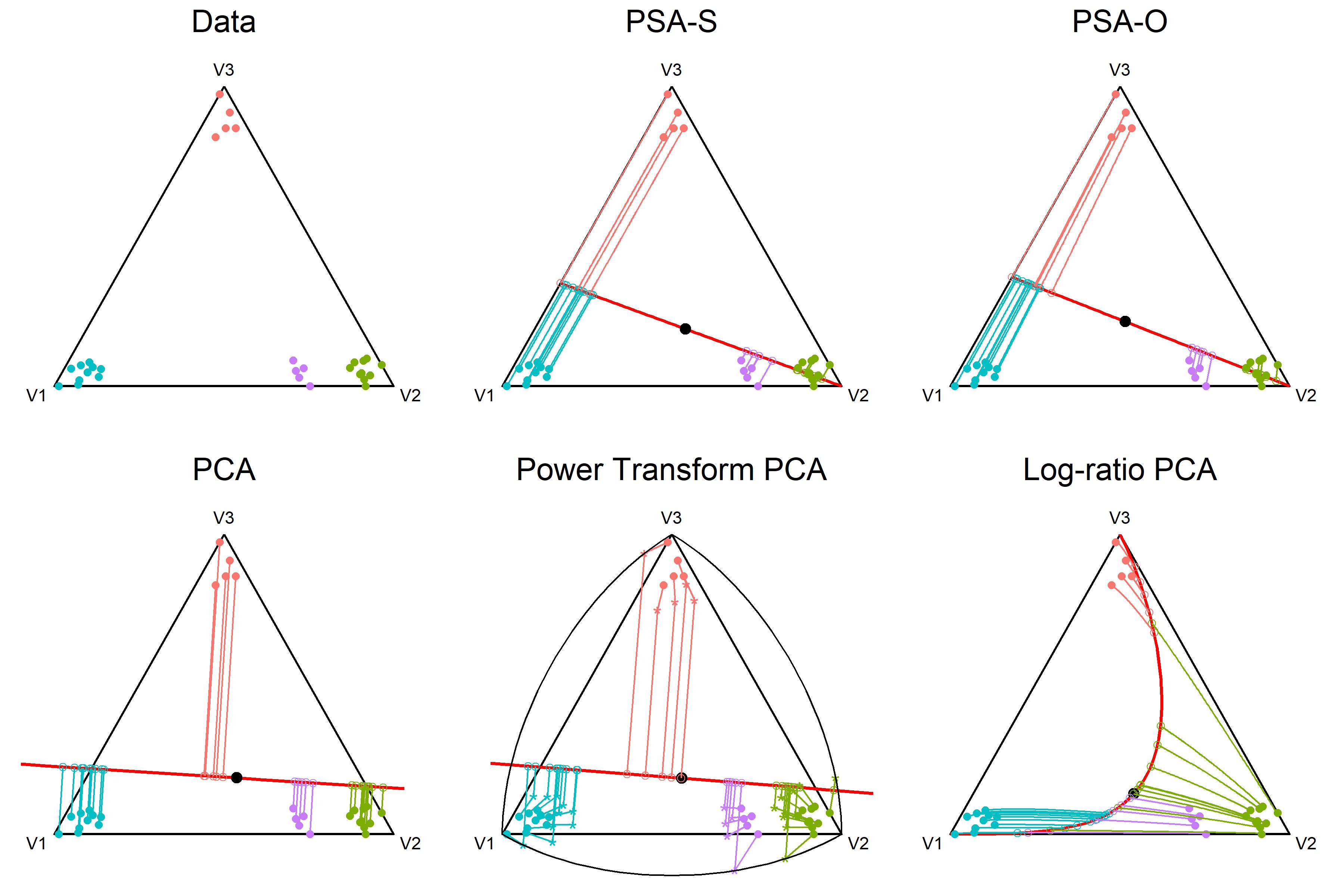}
    \caption{Ternary plots and lower dimensional approximations for Example 1. In each ternary plot represented are the data points (solid circles with different colors representing clusters), the 1-dimensional approximating subset (red line), 1-dimensional approximation (empty circles), and 0-dimensional approximating subset (mean or backwards mean; black solid circle). Shortest-path residuals connect each data point to its 1-dimensional approximation.
    For power transform PCA, the data points after the power transform are shown orthogonally projected onto the hyperplane spanned by the simplex (asterisks).
    }
    \label{fig:ex1_ternary_pc}
\end{figure}

The first simulated data set is 2-dimensional, shown using a ternary plot in the top left panel of Figure \ref{fig:ex1_ternary_pc}. The data set consists of four clusters with centers $\boldsymbol{p}_1 = (0.05, 0.05, 0.9)$, $\boldsymbol{p}_2 = (0.05, 0.9, 0.05)$, $\boldsymbol{p}_3 = (0.9, 0.05, 0.05)$, and $\boldsymbol{p}_4 = (0.25, 0.7, 0.05)$. Each cluster contains either 5 or 10 data points that are randomly drawn from around the cluster centers following an isotropic Gaussian distribution with standard deviation of $0.04$. To ensure the vectors lie on the unit 2-simplex, negative entries were reset to zero and the resulting vectors were renormalized to have unit sum. Consequently, 5.5\% of the entries of the data set were zero. The unbalanced size of the clusters is intended to contrast PSA-S and PSA-O.

Figure \ref{fig:ex1_ternary_pc} illustrates lower dimensional representations produced by the methods. 
The rank 1 approximating subsets (red line segments) demonstrate the fact that the approximating subsets of PSA are subsimplices. In addition, the first modes of variation shown as red lines are linear in $\Delta_d$. PCA also provides linear modes of variation, however the rank 1 approximations of PCA leave the simplex, which is a substantial drawback of PCA. The power transform PCA plot suggests that power transformation distorts the original data and again the approximations leave the simplex. Lastly, log-ratio PCA provides compositional lower dimensional representation but the log transform results in curved modes of variation when mapped back onto the simplex. Note that in the first mode of variation of log-ratio PCA, the green points are so spread that they are not a clear cluster.

\begin{figure}[t!]
    \centering
    \begin{subfigure}{\textwidth}
        \centering
        \includegraphics[width = 0.7\textwidth]{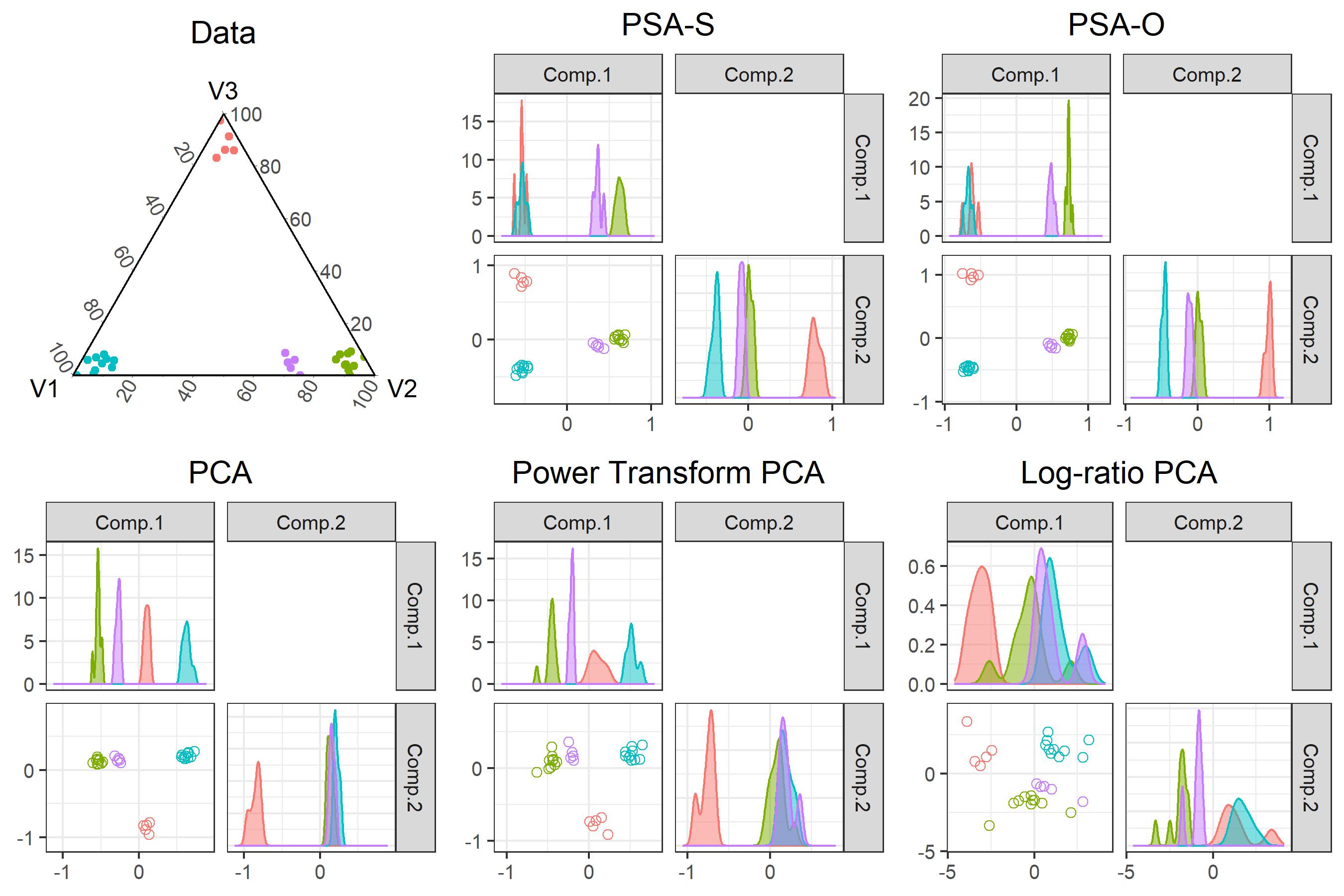}
        \caption{Scores scatter plot matrices for simulated data}
    \end{subfigure}
    \begin{subfigure}{\textwidth}
        \centering
        \includegraphics[width=\textwidth]{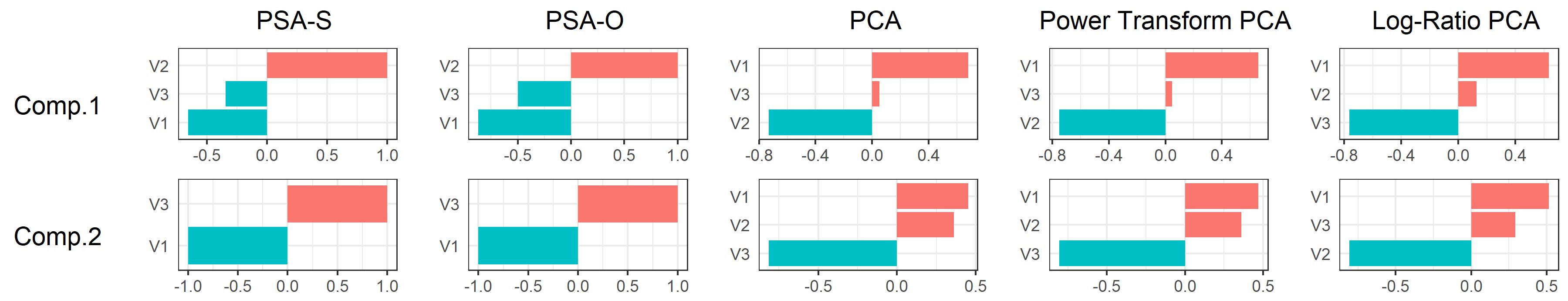}
        \caption{Loading plots for simulated data}
    \end{subfigure}
    \caption{The scores and the loadings for simulated data. The first mode of variation of PSA-S and PSA-O is V2 versus the others, and the second mode of variation occurs between V1 and V3. Separation between clusters are obscured in log-ratio PCA scores.}
    \label{fig:ex1_score}
\end{figure}

The scores scatter plot matrices are shown in Figure \ref{fig:ex1_score} (a). The off-diagonal panels are the scatter plots of the scores and the diagonal panels are the density plots for the scores of each rank colored by cluster. In each of the matrices, the same axes were used for both x-axis and y-axis and for all panels to enhance comparison of scales. The corresponding loading vectors (reflecting the impact of each feature on the mode of variation) are shown in Figure \ref{fig:ex1_score} (b). 

The first column of Figure \ref{fig:ex1_score} (b) suggests that the first mode of variation for PSA-S occurs between V2 and a combination of V1 and V3. The corresponding panel of Figure \ref{fig:ex1_score} (a) confirms that a large positive first score is associated with high V2 content (green and purple clusters) and a large negative first score is associated with high V1 or V3 content (red and cyan clusters). On the other hand, the lower panel of the loading plots suggests that the second mode of variation occurs between V1 and V3 as confirmed in the scores plot. PSA-O provides a similar result in this example.
The PCA panels of Figure \ref{fig:ex1_score} allows a similar interpretation but provides a rotated view with a different emphasis of the clusters. The power transform PCA shows a similar result to PCA with more noisy scores. 
Interpretation of the scores of log-ratio PCA is more involved due to curved modes of variation. A large positive first score is associated with the points near the V1 (cyan cluster) and a large negative value first score is associated with the points near V3 (red cluster). However, a medium level of the first score is associated with the point near V2 (green and purple clusters). We remark that this unexpected association does not happen for the other methods. In addition, the clear separation of the clusters is significantly blurred in the log-ratio scores plot.

\begin{figure}[t!]
    \centering
    \includegraphics[width = 0.8\textwidth]{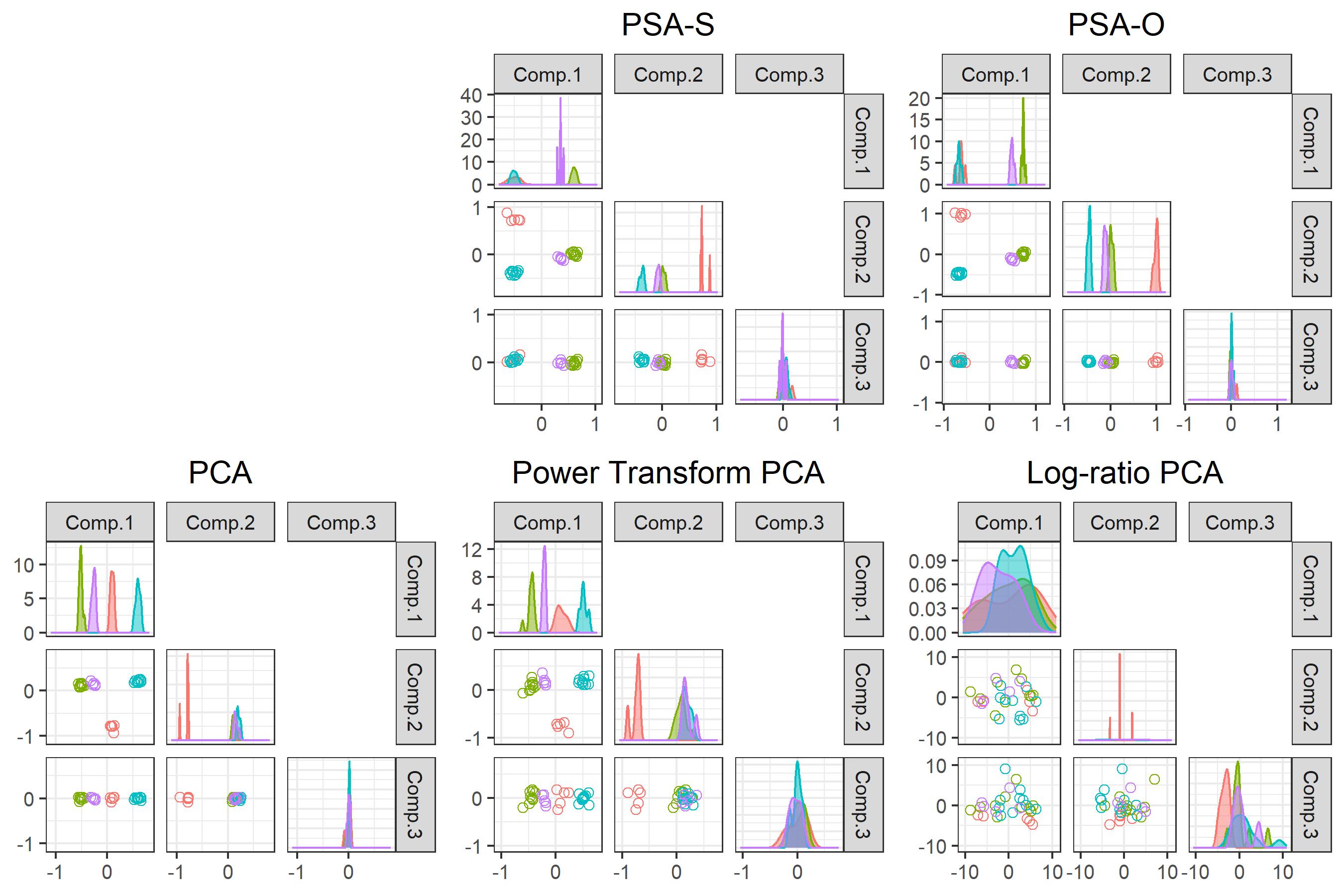}
    \caption{Scores scatter plot matrices for simulated data with pure noise variables. PSA-S, PSA-O, PCA, and the power transform PCA are resilient against the existence of added noise variables while the log-ratio PCA was strongly affected.}
    \label{fig:ex2_score}
\end{figure}

To compare how the five different methods handle low-proportion pure noise variables, we concatenated each vector of simulated data with three random numbers that were independently drawn from $\mathcal{N}(0,0.04^2)$. Again, negative values were reset to zero and renormalized to have unit sum. Figure \ref{fig:ex2_score} shows the first three scores of the methods. The distribution of the first two scores of PSA-S, PSA-O, PCA, and power transform PCA are similar to those in Figure \ref{fig:ex1_score}, implying that these methods are resilient to the existence of added noise variables. The scores of the log-ratio PCA do not show strong differences between the clusters, suggesting that the first three modes of variation for log-ratio PCA were sensitive to the pure noise variables.

In summary, PSA-S and PSA-O provide clearer visual separation of this type of cluster that lives near the edge of the simplex than power transform PCA or log-ratio PCA. Furthermore, PSA-S and PSA-O are more resistant to pure noise variables. Conventional PCA gives similar performance, but at the cost of low dimensional approximations that are not in a subsimplex, and thus no longer compositions.

\section{Application to Relative Abundance of Diatom Species} \label{sec:real data}
\citet{taylor2018polar} studied historic relative abundance of diatom species to explore Southern Ocean conditions during the late Pliocene, which had global temperatures similar to what may occur with global warming. 
The diatoms were observed between 89.84mbsf and 56.53mbsf (meters below seafloor) in a drill core from the Antarctic margins.
Across 71 different depth samples, 61 different diatom taxa were identified. The dataset is available at https://agupubs.onlinelibrary.wiley.com/doi/full/10.1002/2017PA003225 as Table S3.

\begin{figure}[t]
    \centering
    \includegraphics[width = \textwidth]{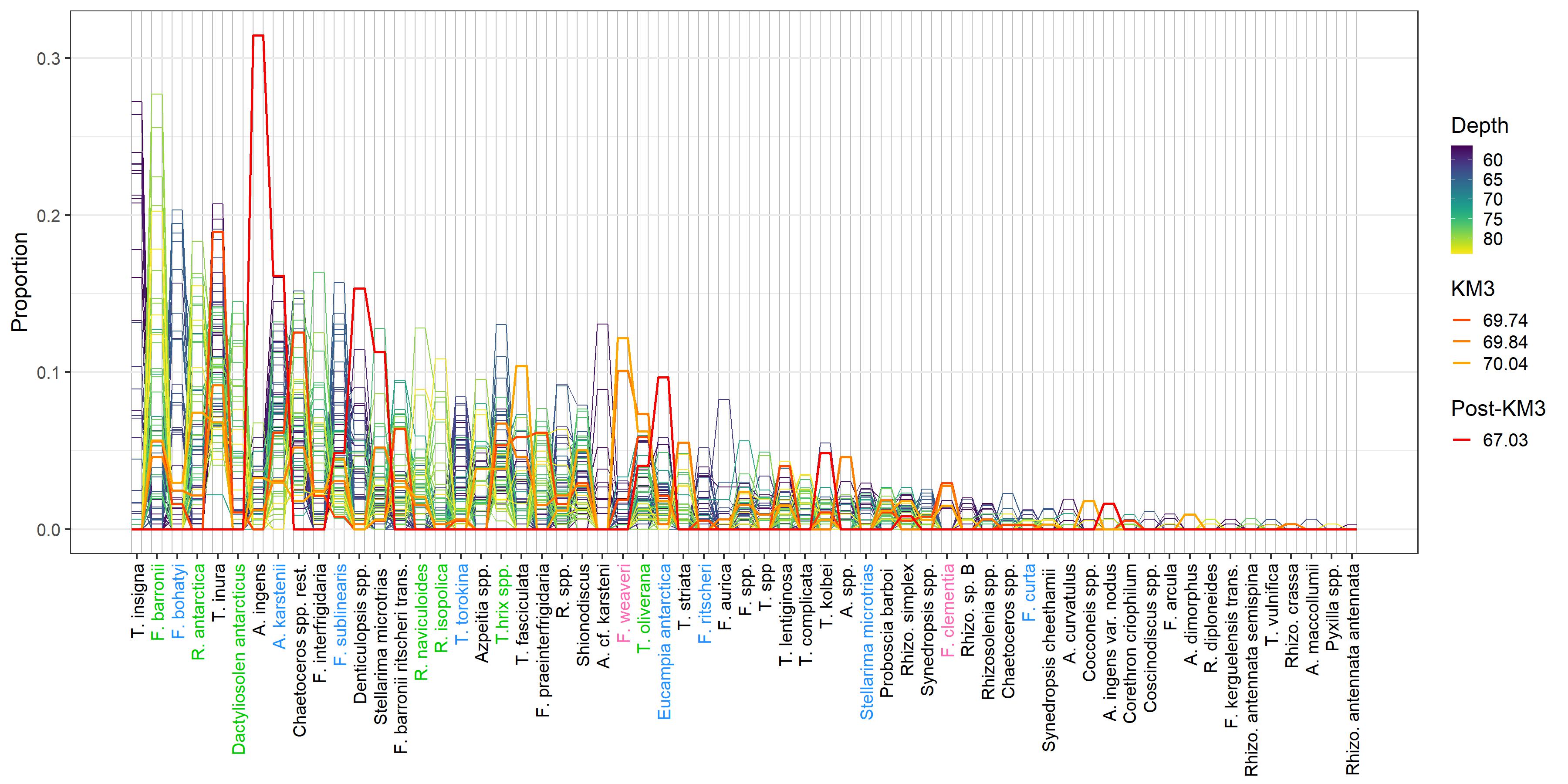}
    \caption{Parallel coordinate plot for the diatom species relative abundance data. Diatom taxa are sorted by decreasing variance, where select ocean temperature-related taxa are colored pink (warm-water), green (ice-tolerant open ocean), or blue (sea-ice affinity) \citep[Fig.~3]{taylor2018polar}. The diatom data set contains a large proportion of zero relative abundances (46\% of entries).}
    \label{fig:diatom distribution}
\end{figure}

The distribution of the diatom species is shown in a parallel coordinate plot (Figure \ref{fig:diatom distribution}) with colors representing the depth of the sample. The diatom dataset contains a large proportion of zero relative abundances (46\% of entries). For the log ratio PCA only, zeros have been replaced by half of 0.0026, the overall minimum nonzero value of the data set and then each measurement was renormalized to have unit sum.

\citet{taylor2018polar} noticed the following major features of the compositions:
(1) starting from the deepest measurements, the relative abundance of sea-ice preferring diatoms increased until
(2) a very short warm period called the KM3 marine isotope stage (70.34mbsf - 69.44mbsf) that contained a high abundance of warm-water diatoms, and
(3) after KM3, the proportion of warm-water diatoms plummeted, there was a high proportion of sea ice preferring diatoms and some open-ocean taxa progressively declined. 
A measurement with extremely low total abundance of diatoms \citep{taylor2018polar} forms a clear outlier at directly after the KM3 period (post-KM3, 67.03mbsf) with extremely high proportion of A.~ingens.
The high abundance of warm-water diatoms during KM3 suggested a dramatic difference in ocean currents and because the KM3 stage coincided with a temporary elevation of atmospheric CO$_2$ concentration to post-industrial levels, \citet{taylor2018polar} inferred that a temporary elevation in atmospheric CO$_2$ was sufficient to trigger a substantial climate response.


\begin{figure}[t!]
    \centering
    \includegraphics[width=0.7\textwidth]{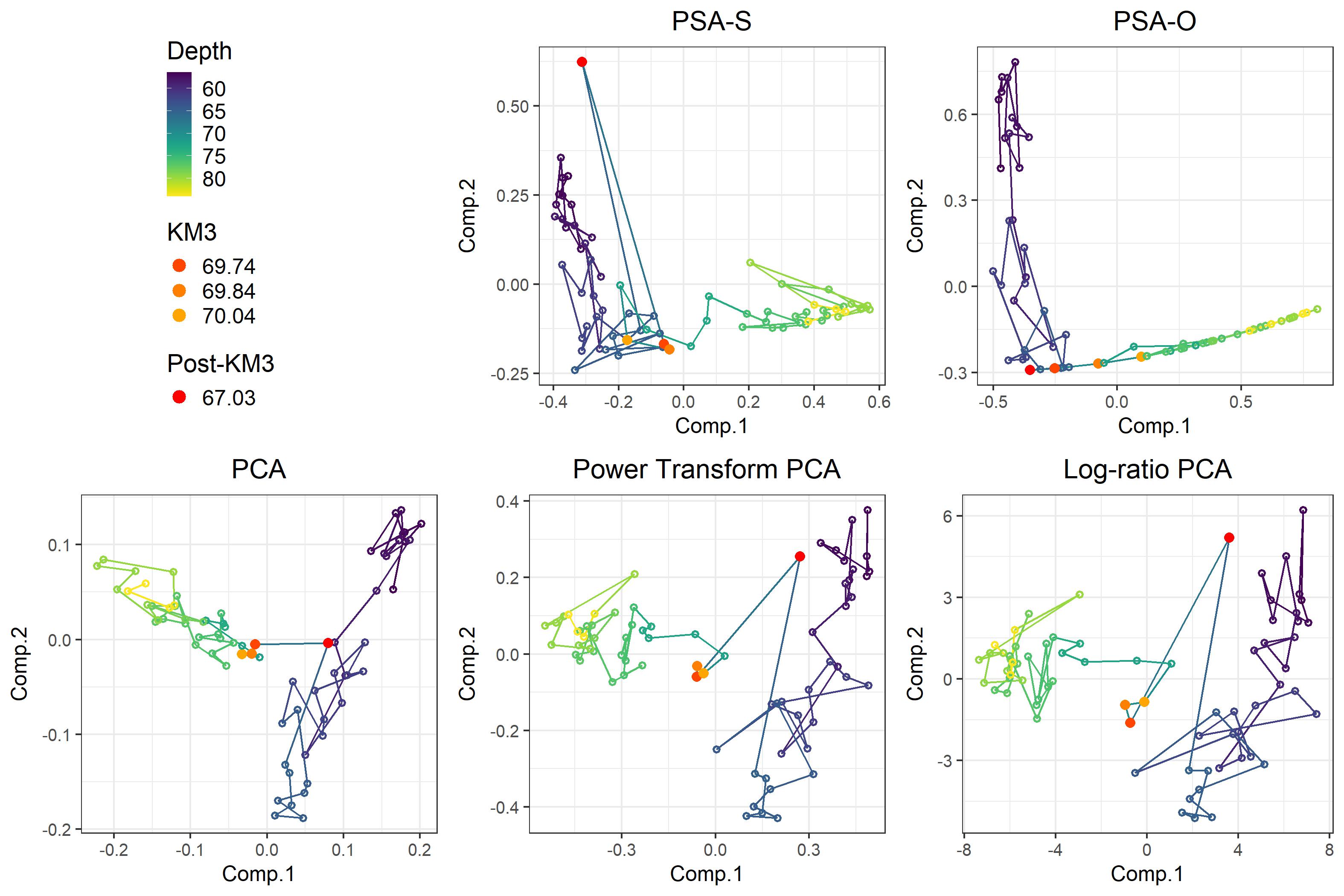}
    \caption{The first two scores of the five methods applied to the diatom relative abundance data. Data points are connected by line segments in depth order. PSA-O scores manifest two arms parallel to each axis with KM3 as a transition point. PSA-S scores show a similar distribution but the second mode of variation is confounded with the post-KM3. Score plots from PCA through the power-transform PCA to the log-ratio PCA show a continuum with added noise.}
    \label{fig:diatom score}
\end{figure}

Figure \ref{fig:diatom score} shows the scatter plots of the first two scores for each of the five methods. Color change along each of the x-axes suggests that all first modes of variation are associated with depth. The unusual nature of the KM3 compositions was not highlighted by any of the methods, however all but PSA-S indicate two distinct phases separated by KM3. Notably, the PSA-O scores are aligned into arms that are close to parallel to each axis and separated by the KM3 period, which provides the clearest and most interpretable modes of variation of all five methods.
The PSA-S scores are similar to those of PSA-O but the transition between two phases does not align with KM3, and the second mode of variation is confounded with the post-KM3 outlier. Indeed, the loading vectors of PSA-S shown in the first row of Figure S3 suggests that while the first loading vector of PSA-S is similar to that of PSA-O, the second loading vector of PSA-S is a mixture of the second and the fourth loading vectors of PSA-O. 
The scores scatter plot of PCA in Figure \ref{fig:diatom score} shows a similar pattern to PSA-O but is rotated compared to the axes and scores appear to contain more noise, making the interpretation of scores in connections with loadings more difficult. The scores plots of the power transform PCA and the log-ratio PCA in Figure \ref{fig:diatom score} are similar to that of PCA but appear even more corrupted by noise, especially for the log-ratio PCA.

The following discussion on the scores plot with a particular emphasis on PSA-O illustrates the strong interpretability of modes of variation and lower dimensional representations obtained from PSA.

\begin{figure}[t!]
    \centering
    \begin{subfigure}{\textwidth}
        \centering
        \includegraphics[width=0.95\textwidth]{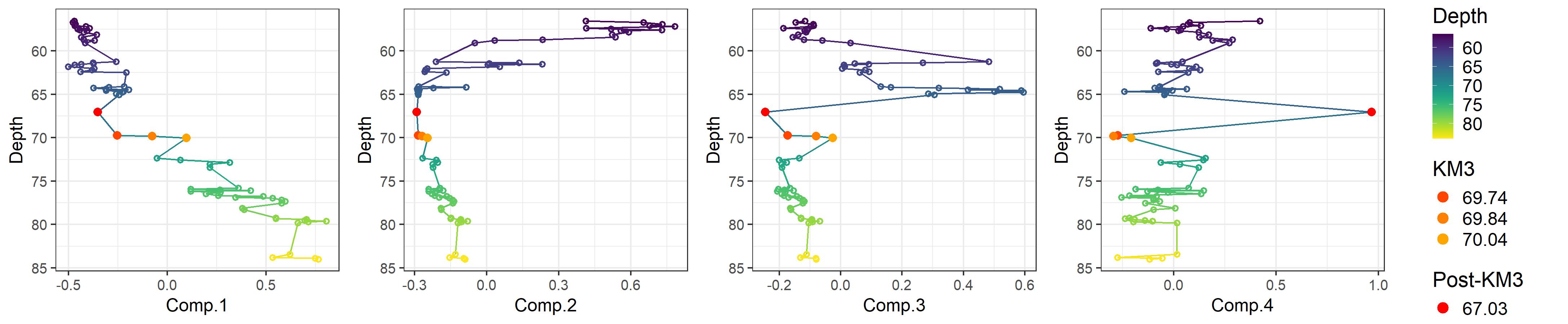}
        \caption{The first four PSA-O scores against depth.}
    \end{subfigure}
    \begin{subfigure}{\textwidth}
        \includegraphics[width=0.95\textwidth]{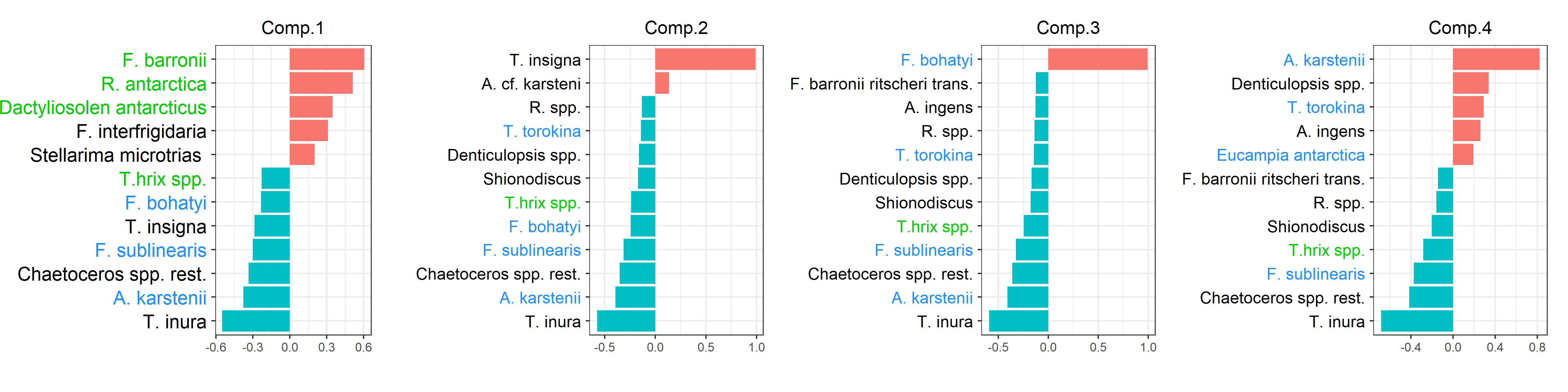}
        \caption{The first four PSA-O loading vectors. The 12 species with the largest absolute value were displayed when there were more than 12 species to show. Some ocean temperature-related taxa are colored pink (warm-water), green (ice-tolerant open ocean), or blue (sea-ice affinity) \citep[Fig.~3]{taylor2018polar}.}    
    \end{subfigure}
    \caption{PSA-O scores and loading vectors for the diatom species relative abundance data.}
    \label{fig:diatom psao}
\end{figure}

Up until the end of KM3 the compositions moved approximately according to the first mode of variation of PSA-O. 
This first mode of variation has large positive loading (Figure \ref{fig:diatom psao}) for several open-ocean diatoms (F. barronii, R. antarctica, D. antarcticus, R. naviculoides), large negative loading for two sea-ice affinity diatoms (F. sublinearis, A. karstenii), and large loadings for a number of other taxa. As the scores of the first mode of variation are decreasing over time, this loading means that there was increasing sea-ice. This confirms observation (1) above by \citet{taylor2018polar}, although they also noted this change towards more sea ice was happening concurrently with changes to a number of other taxa unrelated to ocean temperatures.
After KM3 the compositions maintained very low scores for the first mode of variation, which is consistent with their observation (3), although progressive declines in open-ocean taxa were not highlighted by the PSA-O scores or loadings.

The second mode of variation for PSA-O, shown in the second panel of Figure \ref{fig:diatom psao} (b), has a large positive loading for T. insigna (no particular association with ocean temperatures) and only small loadings for the other taxa, which suggests that after KM3, the largest changes in diatom compositions corresponded to increases with T. insigna over time.
The scores for the third mode of variation for PSA-O jump from negative to large positive values after KM3 and the outlier, and then decreases over time. This third mode has a large positive loading for sea-ice affinity F. bohatyi diatoms, however there are also several much smaller negative loadings for sea-ice affinity. Combined with the second mode of variation, this suggests that there was a rapid jump in F. bohatyi after KM3 and the outlier, and then F. bohatyi had a progressive decline and T. insigna had a progressive increase.
The scores for the fourth mode of variation highlight the outlier at 67.03mbsf and the loadings suggest that the outlier had unusually low proportions of T. inura and Chaetoceros spp, and unusually high proportions of A. karstenii and Denticulopsis spp. Both are consistent with Figure \ref{fig:diatom distribution}, with the unusually low proportions difficult to notice there. The loadings further suggest the outlier has unusually high proportions of T. torokina, A. ingens, and E. antarctica, which broadly aligns with Figure \ref{fig:diatom distribution}. 

\begin{figure}[t!]
    \centering
    \begin{subfigure}{\textwidth}
        \centering
        \includegraphics[width = 0.6\textwidth]{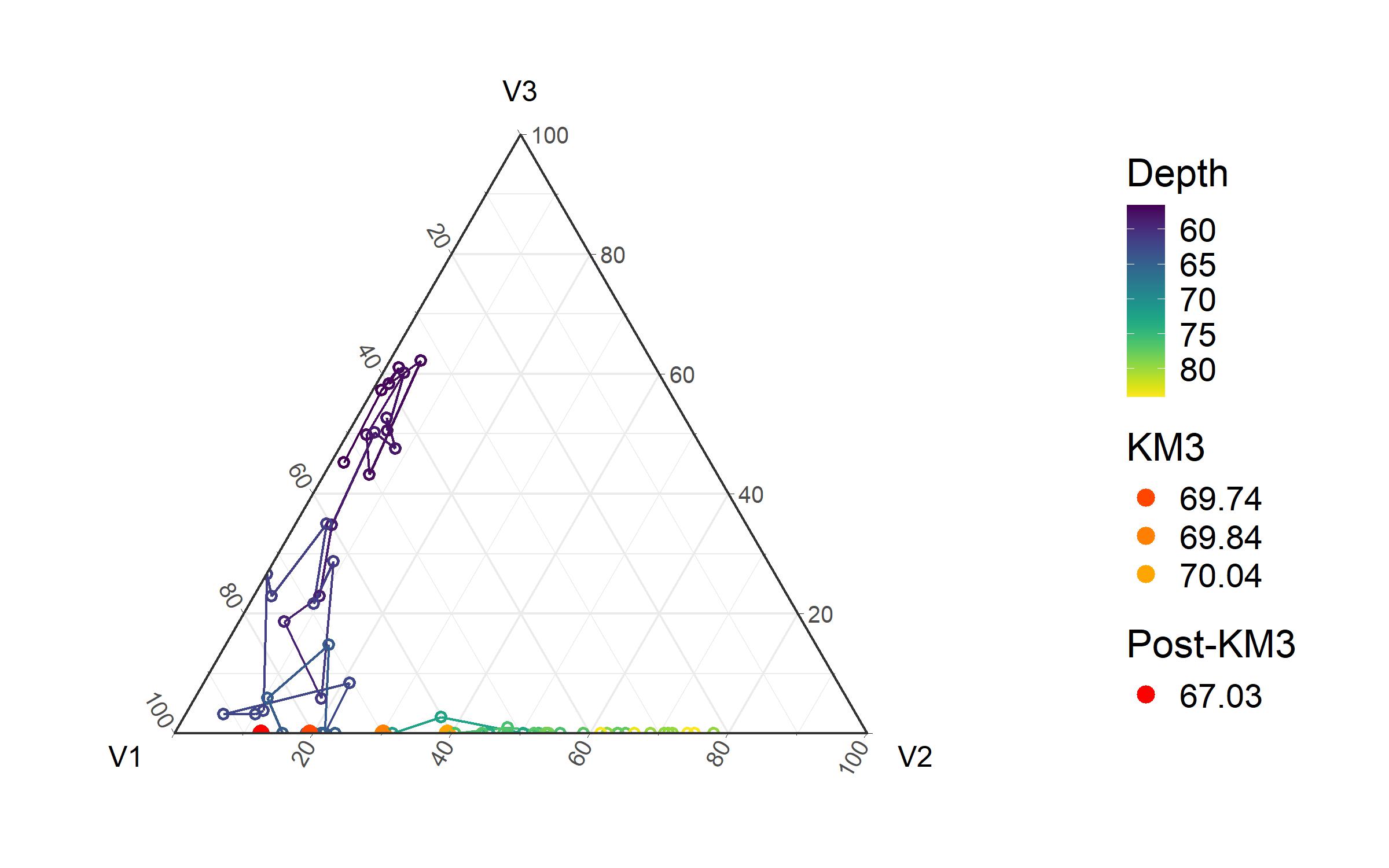}
    \end{subfigure}
    \begin{subfigure}{0.8\textwidth}
    \centering
        \includegraphics[trim = {0 1cm 0 0}, clip, width = 0.95\textwidth]{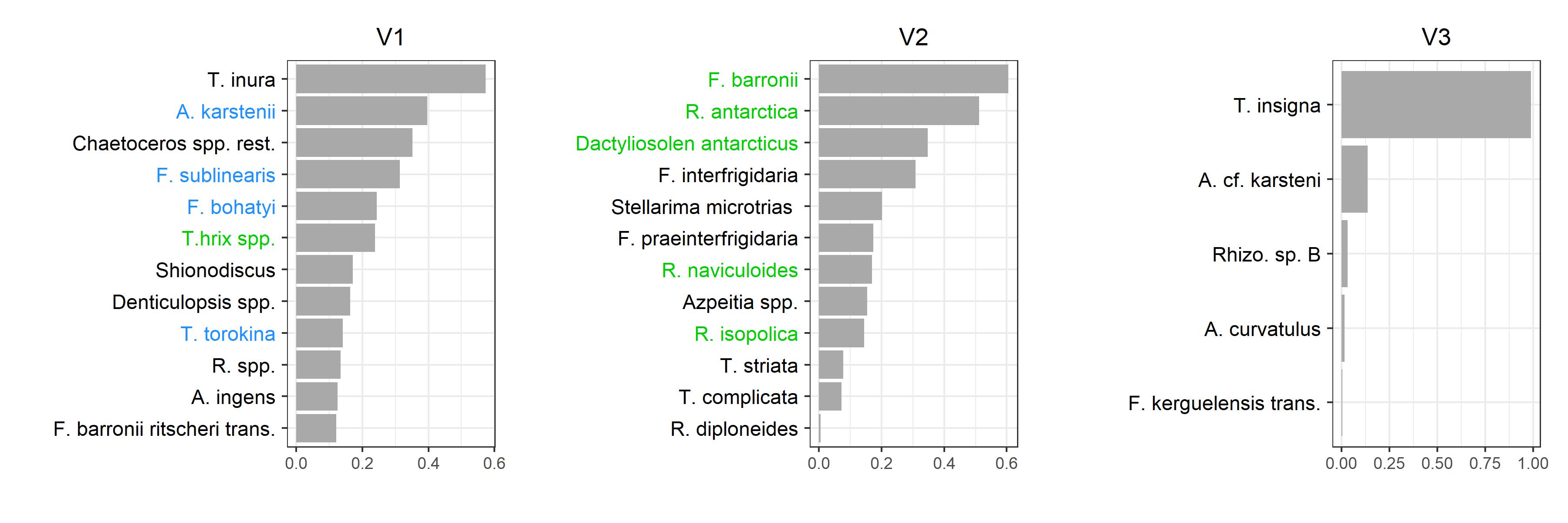}
    \end{subfigure}
    \caption{Rank 2 approximation of PSA-O for the diatom species relative abundance data. The first two vertices show clear association with sea-ice affinity diatoms (blue) and open-ocean diatoms (green).}
    \label{fig:diatom ternary}
\end{figure}

Lastly, the compositional rank 2 representation produced by PSA-O has an effective visualization as a ternary plot (Figure \ref{fig:diatom ternary}).
More directly apparent than in the scores and loading plots (Figures \ref{fig:diatom score}  and \ref{fig:diatom psao}) and \citep[Fig.~3]{taylor2018polar} is the central role of T.~insigna (the major part of vertex V3) with nearly zero relative abundance prior to KM3 and dominating behavior of the compositions after KM3. The trend over time from open-ocean diatoms (largely represented by V2) to sea-ice affinity diatoms (largely represented by V1) is also clear.

\section{Conclusion}

This article proposed a new approach for decomposition of compositional data, applying a backwards PCA framework. At each rank $r$, the proposed optimization can be effectively performed by either closed form expression or a one-dimensional grid search for each of the $r (r+1)/2$ pairs of vertices. The proposed approach provides lower dimensional representations that are compositions and the corresponding modes of variation have compositional interpretations in the sense that the proportion of one group of species increases while that of another group of species decreases. Furthermore, the modes of variation are linear. These properties were illustrated via simulated data sets and analysis of diatom species relative abundance, for which our orthant-based, PSA-O, provided the clearest modes of variation and an informative rank 2 approximation.

\bibliography{main}
\bibliographystyle{apalike}

\setcounter{section}{0}
\setcounter{figure}{0}
\renewcommand{\thesection}{S\arabic{section}}
\renewcommand{\thefigure}{S\arabic{figure}}

\clearpage
{\centering \LARGE Supplementary of Principal Subsimplex Analysis \par}
\section{Backwards Principal Component Analysis}\label{secS:backwards PCA}

For a data set lying on a manifold, backwards PCA searches a nested sequence of submanifolds of decreasing dimensions that fit the data. These submanifolds provide lower dimensional approximations. Moreover, the differences of rank $r+1$ and rank $r$ approximations will be a key component for modes of variation of backwards PCA. (See Section \ref{secS:modes of variation} for more discussion on modes of variation.)

Specifically, backwards PCA starts from the full rank ($d$) representation of data $\mathcal{S}_d$ and finds a $(d-1)$-dimensional submanifold $\mathcal{S}_{d-1}$. One identifies a $(d-2)$-dimensional submanifold $\mathcal{S}_{d-2}$ within $\mathcal{S}_{d-1}$ and continue this procedure until reaching a 0-dimensional submanifold. Each $r$-dimensional submanifold is referred to as the rank $r$ approximating subset of the data. In particular, the 0-dimensional submanifold is called the \textit{backwards mean}, which is frequently a single point though it could be a discrete collection of points in general. 

Principal Nested Spheres (PNS) is an application of backwards PCA to spherical data, which is the first concrete implementation of backwards PCA. Preceding the development of PNS, important precursors include Principal Geodesic Analysis \citep{fletcher2004principal} and Geodesic Principal Component Analysis (\citealt{huckemann2006principal}). These precursors find a sequence of submanifolds spanned by the increasing number of geodesics. While geodesic based approaches have proved its efficacy in many applications, their utility is sometimes constrained by inherent limitations. For example, in spheres, submanifolds spanned by geodesics are the great spheres --the intersections of the sphere with hyperplanes passing through the origin-- thus geodesic-based approaches cannot explain spherical patterns beyond those distributed along the great circles. PNS was conceived to find small spheres, which are the intersections of the sphere with hyperplanes not necessarily passing through the origin. The procedure starts from a $d$-dimensional unit sphere, finds a $(d-1)$-dimensional small sphere, finds a $(d-2)$-dimensional small sphere within it, and iterate the procedure until it reaches the backwards mean. PNS has been adapted to a variety of spaces which involve spheres in different ways. Examples include the space of skeletal representations (\citealt{pizer2013nested}), polyspheres (\citealt{eltzner2015dimension}), and high-dimensional tori (\citealt{eltzner2018torus}, \citealt{zoubouloglou2023scaled}). 

Another application of backwards PCA to nonnegative matrix factorization (\citealt{zhang2015nested}) illustrates distinction between Backwards PCA and many dimensionality reduction methods in interpretability. In most of the nonnegative matrix factorization algorithms, the rank $r$ approximation has no trivial relationship with the rank $r+1$ approximation. In contrast, in backwards PCA, the rank $r$ approximation of the object space is a subset of the rank $r+1$ approximation thus it is easy to understand the relationship between approximations of different ranks. Moreover, the differences between rank $r$ approximations and rank $r+1$ approximations naturally define modes of variation. 

\section{Modes of Variation by PSA}\label{secS:modes of variation}

Modes of variation are the important `directions' in which the data vary. In the Euclidean Principal Component Analysis context, the loading vectors (equivalently principal components) are interpreted as the modes of variation. Modes of variation are not necessarily linear, as in the case of Principal Geodesic Analysis where modes of variation are geodesics passing through the Frechet mean.

More generally, \citealp{marron2021object} defined a mode of variation as a one parameter family of objects in the data space. The best-known modes of variation are those for Euclidean PCA which are orthogonal straight lines passing through the mean of the data. Thinking of a mode of variation as a family of data objects effectively generalizes the definition of modes of variation for Euclidean PCA to nonlinear spaces. For example, the modes of variation for principal geodesic analysis fit this definition as a geodesic is a one-parameter family of points. 

A natural choice of the first mode of variation for backwards PCA is the collection of one-dimensional approximations. For example, in PNS, the first mode of variation was defined as the rank 1 approximating circle. However, the higher order modes of variation for backwards PCA have not yet been formally defined. In this paper, we propose the higher order modes of variation for PSA as follows.

\begin{figure}[H]
    \centering
    \begin{subfigure}{0.3\textwidth}
        \centering
        \includegraphics[trim = {0 0 16.5cm 0}, clip, width = \textwidth]{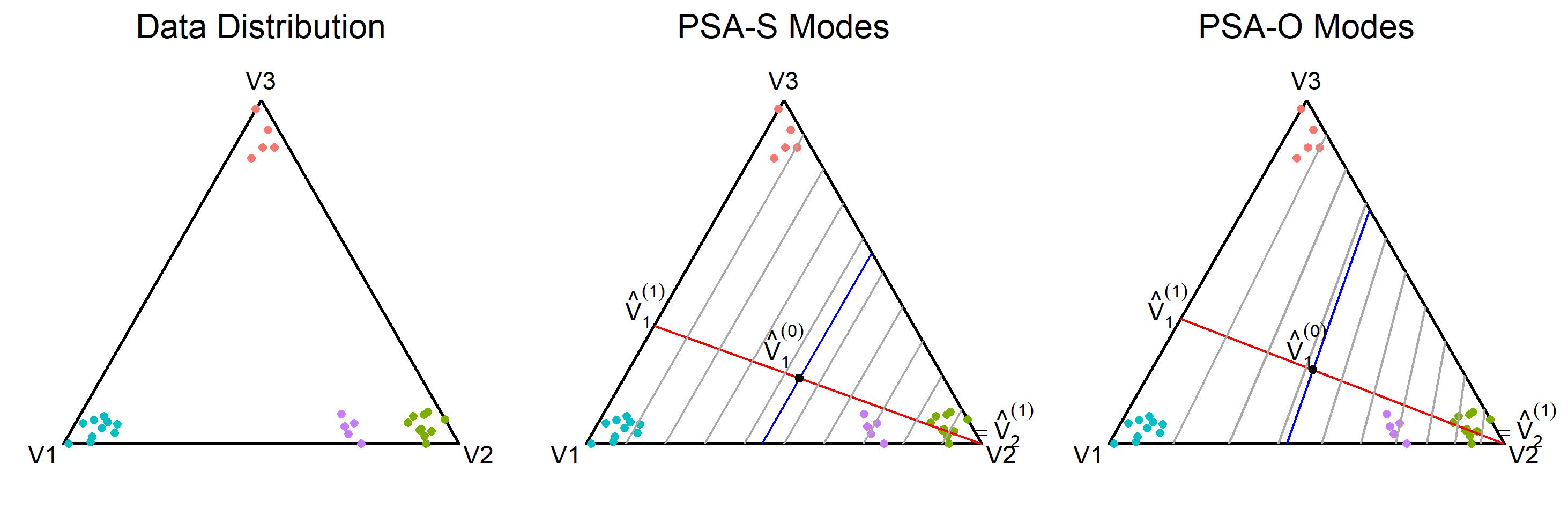}
        \caption{}
    \end{subfigure}
        \begin{subfigure}{0.3\textwidth}
        \centering
        \includegraphics[trim = {8.25cm 0 8.25cm 0}, clip, width = \textwidth]{figures/Examples/ex1_PSA_modes.jpeg}
        \caption{}
    \end{subfigure}
        \begin{subfigure}{0.3\textwidth}
        \centering
        \includegraphics[trim = {16.5cm 0 0 0}, clip, width = \textwidth]{figures/Examples/ex1_PSA_modes.jpeg}
        \caption{}
    \end{subfigure}
    \caption{Application of PSA to Example 1 illustrating modes of variation. (a) Distribution of data. (b), (c) First two modes of variation and rank 2 trajectories of PSA-S and PSA-O. In each of the panels, the red line segment is the first mode of variation and the black solid dot is the backwards mean. The gray line segments show trajectories of equally spaced points on the space of the first mode of variation and the blue line segment shows the second mode of variation.}
    \label{fig:PSA modes}
\end{figure}

The idea behind modes of variation in PSA is illustrated using Example 1 in Section \ref{sec:simulation} of the paper. The middle panel of Figure \ref{fig:PSA modes} shows a different view of the application of PSA-S to Example 1. The plot illustrates that the first merge occurred between the vertices $V_3$ and $V_1$ giving the new vertex $\hat{V}^{(1)}_1=0.66V_1+0.34V_3$. The merge yielded the rank 1-approximating subset which is depicted as the red line segment. The vertex $V_2$ which was not merged was relabeled as $V^{(1)}_2$. In the second merge of PSA-S, $\hat{V}^{(1)}_1$ was merged with $V_2$ to give the backwards mean 
\[
\hat{V}^{(0)}_1=0.556\hat{V}^{(1)}_1+0.444V_2=0.366V_1+0.444V_2+0.19V_3
\]
which is marked by the black solid dot. 

More importantly, the panel also illustrates the way each point in $\Delta_2$ is approximated by PSA-S. Each of the gray line segments is the collection of the points with the same rank 1 approximation. We call these collections the \textit{rank 2 trajectories}. The index rank 2 indicates that the collection is chosen from the rank 2 approximating subset. Generally, a \textit{rank $r$ trajectory} is the one-dimensional subset of the rank $r+1$ approximating subset all of which have the same rank $r$ approximation. The unique rank 1 trajectory is the red line segment, which is the rank 1 approximating subset itself.

We define the $r$th mode of variation as the rank $r$ trajectory of the backwards mean. The first and the second modes of variation are shown in Figure \ref{fig:PSA modes} as the red and blue line segments, respectively. This definition generalizes the modes of variation for PCA and PGA. Moreover, this definition of higher order modes of variation can be naturally extended to backwards PCA on other spaces, including PNS.

The right panel of Figure \ref{fig:PSA modes} is the same display for PSA-O. It is clear from the figures that while the trajectories of PSA-S are parallel to a side of the ternary plot, those of PSA-O are not. This is because in PSA-O, points are projected to the orthant space, then projected onto the great circle that corresponds to the red line segment along great circles that are perpendicular to that great circle, and then projected back to the simplex.

\begin{figure}[t!]
    \centering
    \includegraphics[width=\textwidth]{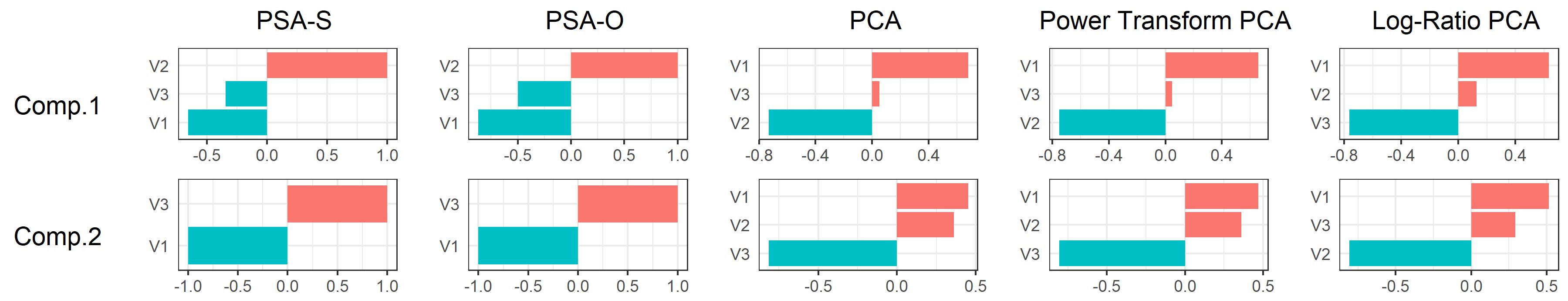}
    \caption{Loading plots for Example 1. The first loading plot of PSA-S indicates that the first mode of variation is $V2$ versus $0.66V_1+0.34V_3$ the same as the observation made in the ternary plot, Figure \ref{fig:PSA modes}.}
    \label{figS:ex1_loading}
\end{figure}

An alternative way of describing a mode of variation, instead of using a set of data objects, is to use the difference of two merged vertices (as vectors in $\mathbb{R}^{d+1}$), which we call a \textit{loading vector}. In the PSA-S panel of Figure \ref{fig:PSA modes}, the direction of the second mode of variation is the difference $V_3-V_1$ of the merged vertices $V_3$ and $V_1$. The direction of the first mode of variation is the difference $V_2-\hat{V}^{(1)}_1$ of the merged vertices $V_2$ and $\hat{V}^{(1)}_1$. This observation leads to the definition of loading vectors in Section \ref{subsec:PSA-S}. We may represent modes of variation of PSA-O in the same way, although the trajectories of PSA-O are not exactly parallel to each other. Figure \ref{figS:ex1_loading} visualizes the loading vectors using bar plots (the same figure as Figure \ref{fig:ex1_score} (b).) A positive score implies high content of the elements with the red bars. The red bars and the green bars add up to 1 and -1, respectively, due to the facts that two merged vertices are in $\Delta_d$ and that two merged vertices consist of disjoint sets of elements.

\begin{figure}[H]
    \centering
    \includegraphics[width = \textwidth]{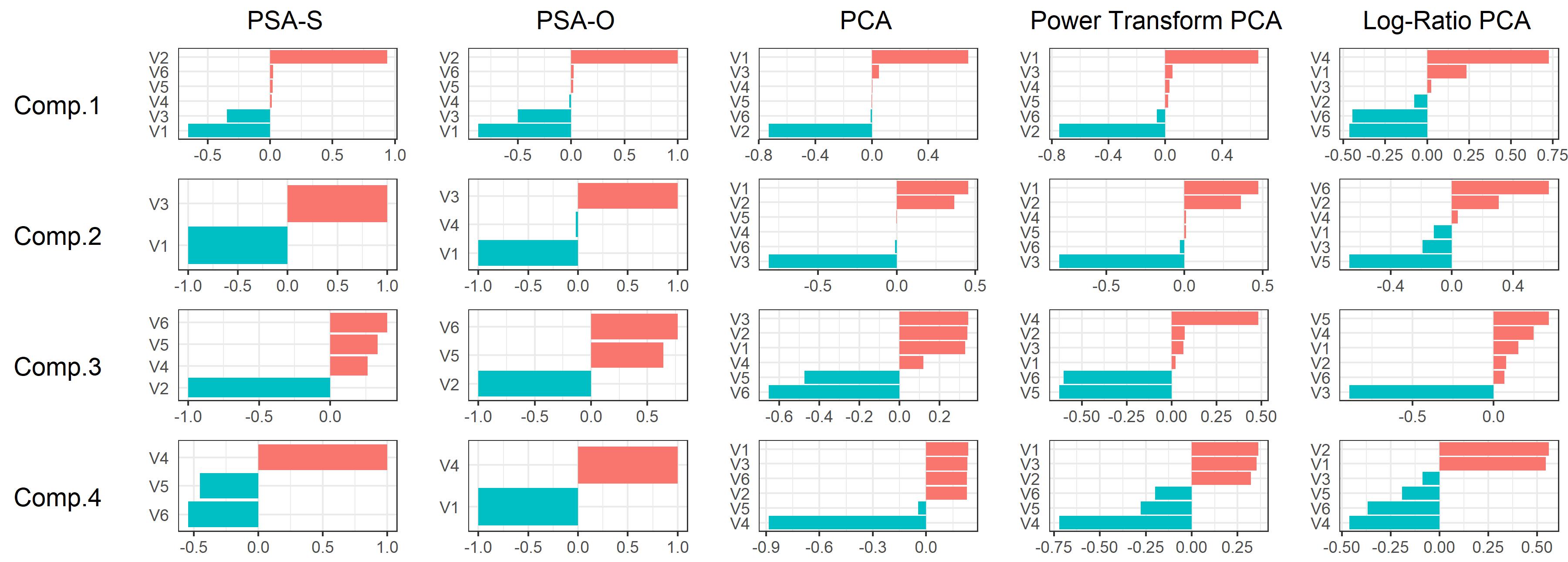}
    \caption{Loading plots for Example 2.}
    \label{figS:ex2_loading}
\end{figure}

Figure \ref{figS:ex2_loading} uses the same bar plots to display the loading vectors of Example 2.

\clearpage
\section{Algorithms of PSA}

\begin{algorithm}[]
\SetAlgoLined
\caption{Principal Subsimplex Analysis via Simplices (PSA-S)}
\label{alg:PSA-S}
Initialize $\boldsymbol{v}^{(d)}_j=\boldsymbol{e}_j$ for $j=1,\cdots,d+1$ and
$\hat{\boldsymbol{x}}^{(d)}_i=\boldsymbol{x}_i$ for $i=1,\cdots,n$.\\
\For{$r = d,\cdots,1$}{
    \For{\text{pair of vertices from } $\boldsymbol{v}^{(r)}_1,\cdots,\boldsymbol{v}^{(r)}_{r+1}$}{
    Merge the pair of vertices at optimal $\hat{\alpha}_r$ and define $\hat{\boldsymbol{v}}^{(r-1)}_1$ and $S_{r-1}$.\\
    Compute $\hat{\boldsymbol{x}}^{(r-1)}_i$, $s^{(r)}_i$, and $\boldsymbol{l}_r$.
}
Find the pair of vertices that minimizes the sum of squared scores.
}
\end{algorithm}

\begin{algorithm}[]
\caption{Principal Subsimplex Analysis via Orthants (PSA-O)}
\label{alg:PSA-O}
Project $\boldsymbol{x}_i$ onto $\mathcal{O}_d$ via $\tilde{\boldsymbol{x}}_i=\boldsymbol{x}_i/\Vert\boldsymbol{x}_i\Vert_2$.\\
Initialize $\tilde{\boldsymbol{v}}^{(d)}_j=\boldsymbol{e}_j$ for $j=1,\cdots,d+1$ and $\tilde{\boldsymbol{x}}^{(d)}_i=\tilde{\boldsymbol{x}}_i$ for $i=1,\cdots,n$.\\
\For{$r = d,\cdots,1$}{
\For{pair of vertices from $\boldsymbol{v}^{(r)}_j$ and weight $\alpha_r\in \textnormal{grid}([0,1])$}{
    Merge the pair of vertices at the ratio to define $\tilde{\boldsymbol{v}}^{(r-1)}_1$ and $O_{r-1}$.\\
    Compute $\tilde{\boldsymbol{x}}^{(r-1)}_i$ and $s^{(r)}_i$.
}
Find the pair of vertices and the weight that minimize the sum of squared scores.
}
For each $r=d,\cdots,0$, project $O_r$ and $\tilde{\boldsymbol{x}}^{(r)}_i$ back onto $\Delta_d$ to get $S_r$ and $\hat{\boldsymbol{x}}^{(r)}_i$.\\
Compute $\boldsymbol{l}_r$ for $r=d,\cdots,0$.
\end{algorithm}

\clearpage
\section{Alternative Approaches to Compositional Data} \label{sec:alternatives}
Let us denote the component-wise arithmetic by $\frac{\boldsymbol{x}}{b}+a=\left(\frac{x_1}{b}+a,\cdots,\frac{x_{d+1}}{b}+a\right)^\top$ where $\boldsymbol{x}\in\mathbb{R}^{d+1}$ and $a,b\in\mathbb{R}$. Similarly, component-wise logarithm and power transformations are denoted by $\log\boldsymbol{x}=(\log x_1,\cdots,\cdots,\log x_{d+1})$ and $\boldsymbol{x}^\alpha = (x_1^\alpha,\cdots,x_{d+1}^\alpha)$ for any $\alpha>0$. Also, $\boldsymbol{x}_{-{(d+1)}}=(x_1,\cdots.x_d)$ where $\boldsymbol{x}=(x_1,\cdots,x_d,x_{d+1})\in\mathbb{R}^{d+1}$. Let $\boldsymbol{1}_p=(1,\cdots,1)\in\mathbb{R}^p$.

\subsection{Log-Ratio Transformations}
The log transform is defined for positive values, so we denote by $\Delta_d^\circ$ the \textit{interior of $\Delta_d$} or the \textit{open unit $d$-simplex}, that is,
\[
\Delta_d^\circ=\left\{\boldsymbol{x}\in\Delta_d, x_j>0, j=1,\cdots,d+1\right\}.
\] \citet{aitchison1986statistical} proposed the \textit{additive log-ratio transformation}, $\boldsymbol{w}_{\text{alr}}:\Delta_d^\circ\to \mathbb{R}^d$ defined by
\begin{equation}
\boldsymbol{w}_{\text{alr}}(\boldsymbol{x}) = \log \left(\frac{\boldsymbol{x}_{-{(d+1)}}}{x_{d+1}}\right).
\end{equation}
Division by the last component gives the additive log-ratio transformation the nice property that the component-wise log transformation $\log \boldsymbol{x}$ does not have: it is a one-to-one mapping from the open unit $d$-simplex to the entire Euclidean space $\mathbb{R}^d$.

This correspondence is particularly useful in parametric analysis of compositional data with no zero entries. A random variable supported on $\Delta_d^\circ$ is said to follow \textit{logistic normal distribution} with parameters $\boldsymbol{\mu}\in\mathbb{R}^d$ and $\boldsymbol{\Sigma}\in\mathbb{R}^{d^2}$, denoted by $\mathcal{L}^d(\boldsymbol{\mu,\Sigma})$, if $\boldsymbol{w}_{alr}(\boldsymbol{x})$ follows $N_d(\boldsymbol{\mu,\Sigma})$. The class of logistic normal distributions is more flexible than the traditional class of Dirichlet distributions in the sense that it can model correlation between components in a way that Dirichlet distributions cannot.

Because the additive log-ratio transformation depends on the choice of ordering of components and it may greatly affect subsequent analyses, \citet{aitchison1986statistical} further proposed the \textit{centered log-ratio transformation}, $\boldsymbol{w}_{\text{clr}}:\Delta_d^\circ\to \mathbb{R}^{d+1}$, which centers each point $\boldsymbol{x}$ by its geometric mean $g(\boldsymbol{x})=\sqrt[d+1]{x_1 \cdots x_{d+1}}$.
\begin{equation}
\boldsymbol{w}_{\text{clr}}(\boldsymbol{x}) = \log \left(\frac{\boldsymbol{x}}{g(\boldsymbol{x})}\right).
\end{equation}
The centered log-ratio transformation is a one-to-one mapping from $\Delta_d^\circ$ to the $d$-dimensional subspace $H=\{\boldsymbol{w}\in\mathbb{R}^{d+1}:\boldsymbol{1}_{d+1}^\top \boldsymbol{w}=0\}$. One can remove the redundant dimension by isometrically identifying the subspace with $\mathbb{R}^d$, which leads to the \textit{isometric log-ratio transformation}, $\boldsymbol{w}_{\text{ilr}}:\Delta_d^\circ\to \mathbb{R}^d$. The identification can be computed by
\begin{equation}
\boldsymbol{w}_{\text{ilr}}(\boldsymbol{x}) = \boldsymbol{H}_{d+1}\boldsymbol{w}_{\text{clr}}(\boldsymbol{x})
\end{equation}
where $\boldsymbol{H}_{d+1}$ is the lower $d\times (d+1)$ submatrix of the Helmert matrix of order $(d+1)$. This refinement is required when a statistical method requires the data to be of full rank.

For principal component analysis purpose, however, there is no real difference between using clr and ilr, but clr transformation is intuitively and computationally simpler, for clr maps all data points into the hyperplane $H$ spanned by the simplex which essentially passes through the mean. Consequently, the last principal component of the clr-transformed data is the normal vector $\boldsymbol{1}_{d+1}$ and variation explained by the principal component is zero. In addition, all lower dimensional approximating subspaces and approximation points of clr-PCA lie on $H$ and these approximations can be effectively mapped back onto the open unit simplex through the inverse the of clr map. 

\begin{figure}[t]
    \centering
    \begin{subfigure}{0.3\textwidth}
        \centering
        \includegraphics[width = \textwidth]{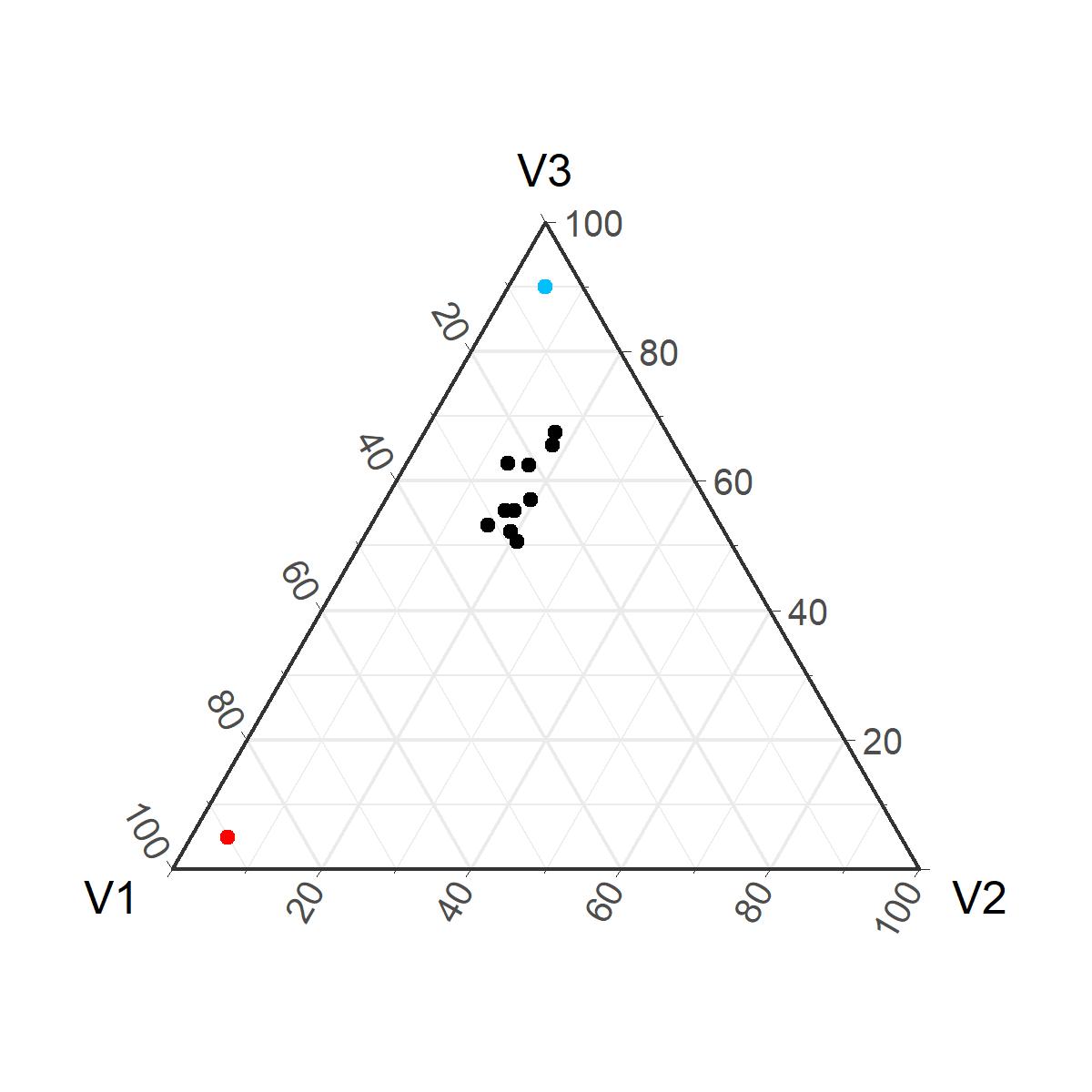}
        \caption{}
    \end{subfigure}
    \begin{subfigure}{0.3\textwidth}
        \centering
        \includegraphics[width = \textwidth]{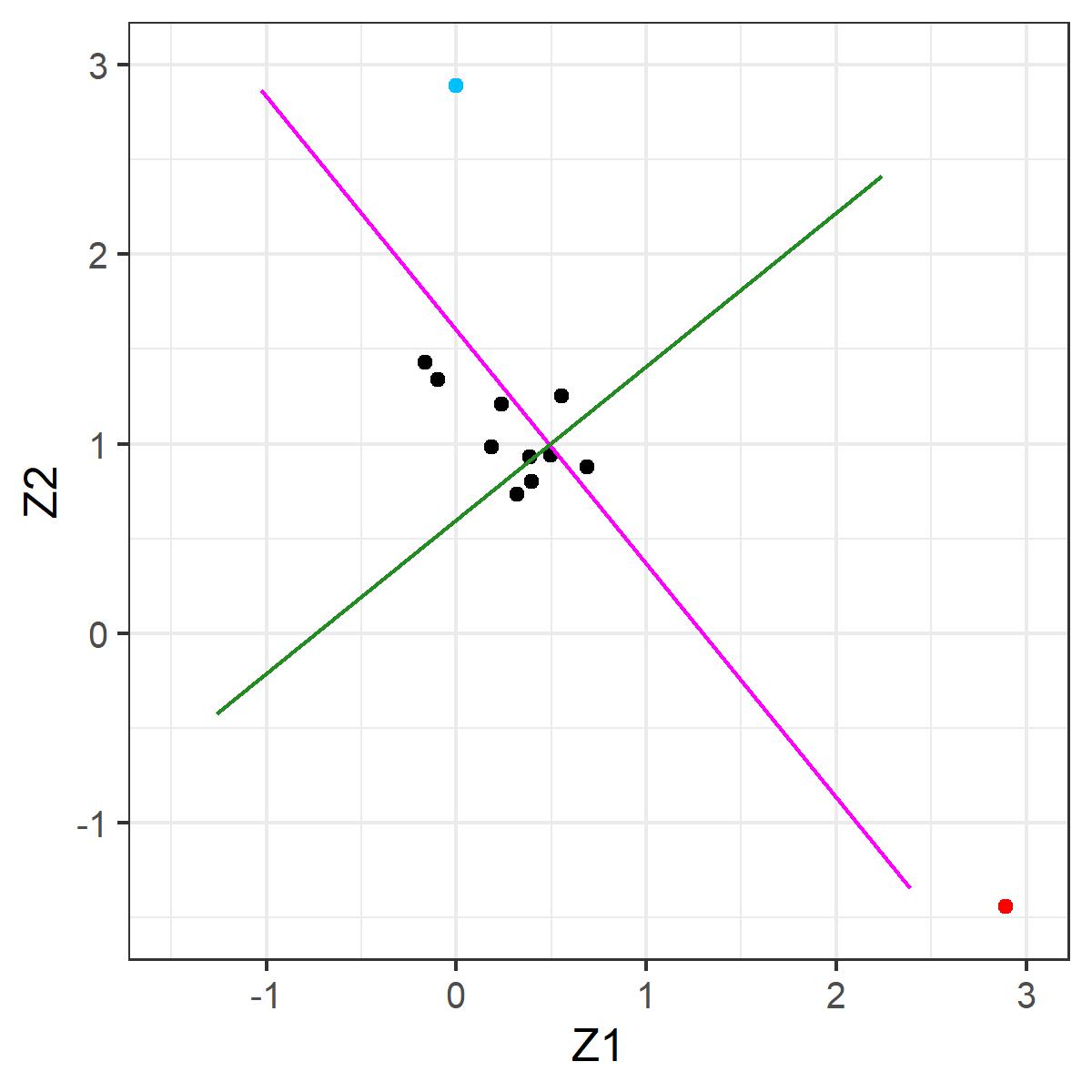}
        \caption{}
    \end{subfigure}
    \begin{subfigure}{0.3\textwidth}
        \centering
        \includegraphics[width = \textwidth]{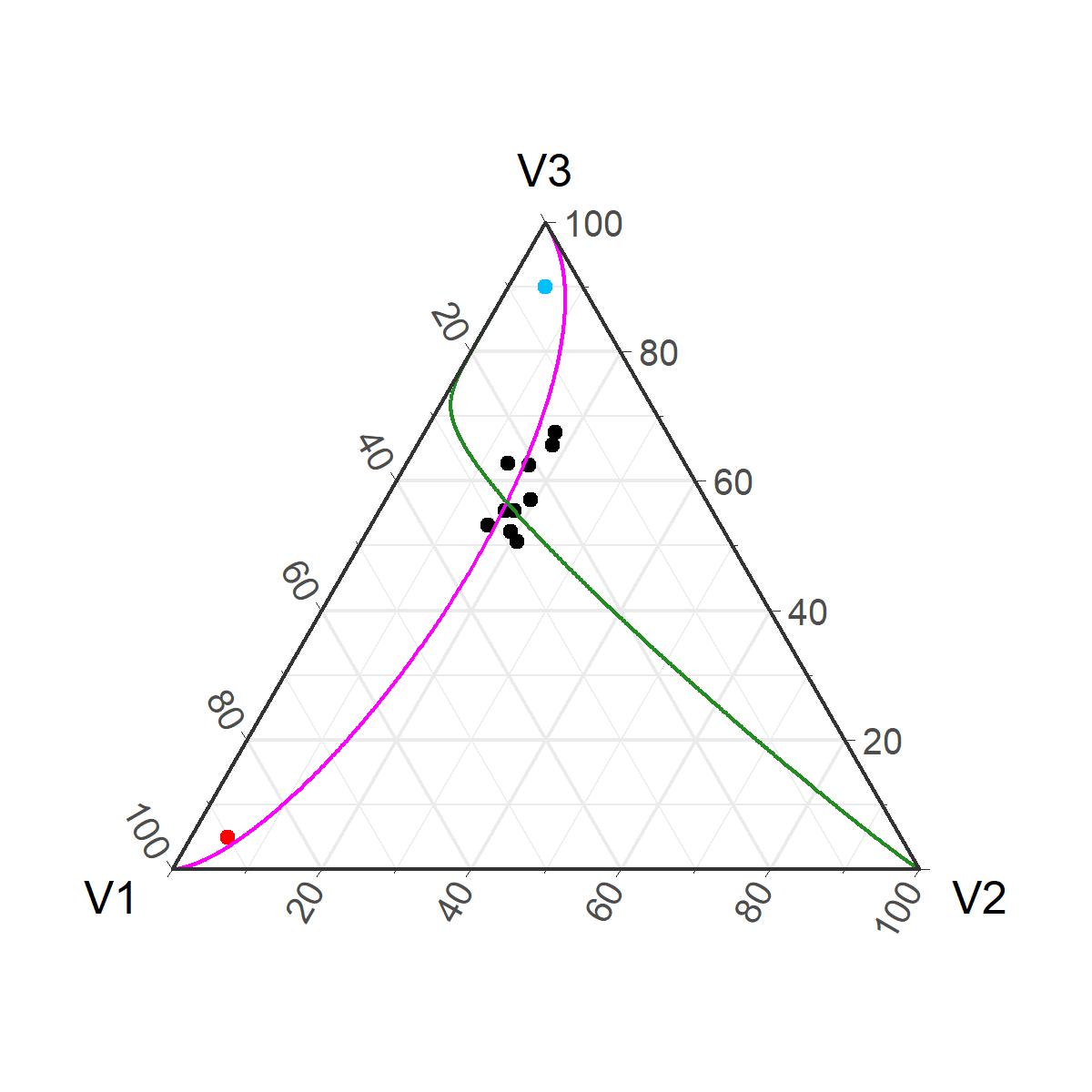}
        \caption{}
    \end{subfigure}    
    \caption{Example of application of the log-ratio PCA to a 2-dimensional compositional data. Two modes of variation exhibit significant curvatures.}
    \label{fig:log-ratio PCA}
\end{figure}

Figure \ref{fig:log-ratio PCA} illustrates an example of 2-dimensional compositional data in which log-ratio PCA is applied. In Figure \ref{fig:log-ratio PCA} (a), two outlying data points are colored blue and red, while the other points are colored black. Figure \ref{fig:log-ratio PCA} (b) shows the distribution of the same data after ilr transformation, along with the magenta and green line segments indicating the first and the second principal directions. These line segments are mapped back onto the simplex in Figure \ref{fig:log-ratio PCA} (c). Note that significant curvature are introduced through the log-ratio transformation. 

The log-ratio transformations are not applicable when a data set contains zeros. Using terminologies in Chapter 11 of \citet{aitchison1986statistical}, zeros in compositional data are called \textit{essential zeros} (or \textit{structural zeros}) if the corresponding parts were truly not there, and \textit{rounded zeros} (or \textit{count zeros}) if the zeros occur because the true values were below detection level. Essential zeros may suggest existence of distinct subpopulations that can be modeled by hierarchical models, for example, zero inflated Poisson models. Poisson component is modeling count zeros, and the point mass at the zero is modeling essential zeros. Rounded zeros are typically replaced by small values through imputation.

\subsection{Power Transformations}

Box-Cox power transformations are also frequently used to handle high skewness, and they generalize the log transformation in the sense that $\lim_{\alpha\to0}\frac{\boldsymbol{x}^\alpha-1}{\alpha}=\log \boldsymbol{x}$. \citet{aitchison1986statistical} proposed to apply Box-Cox power transformation to the ratio of the components, which we denote by $w_{\alpha}^{(1)}:\Delta_d^\circ\to \mathbb{R}^d$,
\begin{equation}
\boldsymbol{w}_{\alpha}^{(1)}(\boldsymbol{x}) = \frac{(\boldsymbol{x}/x_{d+1})^\alpha-1}{\alpha}.
\end{equation}
for some $\alpha>0$. On the other hand, one can consider Box-Cox power transformation on the components but not on the ratio, $\boldsymbol{w}_{\alpha}^{(2)}:\Delta_d^\circ\to \mathbb{R}^{d+1}$,
\begin{equation}
\boldsymbol{w}_{\alpha}^{(2)}(\boldsymbol{x}) = \frac{\boldsymbol{x}^{\alpha}-1}{\alpha}.
\end{equation}
This approach was extended regarding robustness in \citet{scealy2015robust}. In contrast to the case of the centered log-ratio transformation, the data points resulting from power transformations lie on a curved space but not a linear space.


For PCA purposes, $\alpha$ is often chosen so that it maximizes the profile log-likelihood assuming normality after power transformation. Another approach is simply using $\alpha = 1/2$ and consider $\boldsymbol{x}\mapsto \boldsymbol{x}^{1/2}$. Note that this is a translation and a dilation of the Box-Cox power transformation. This transformation effectively maps a simplex onto the nonnegative orthant of the same dimension, which proves particularly advantageous for parametric analysis of compositional data. This enables the utilization of models defined on spheres (\citealt{scealy2011regression}, \citealt{zhu2024spherical}).

\section{Additional Discussion on the Diatom Data}

This section provides additional figures and discussion for the Diatom data.

\subsection{Data Set}

\begin{figure}[t]
    \centering
    \begin{subfigure}{\textwidth}  
        \centering
        \includegraphics[width = 0.8\textwidth]{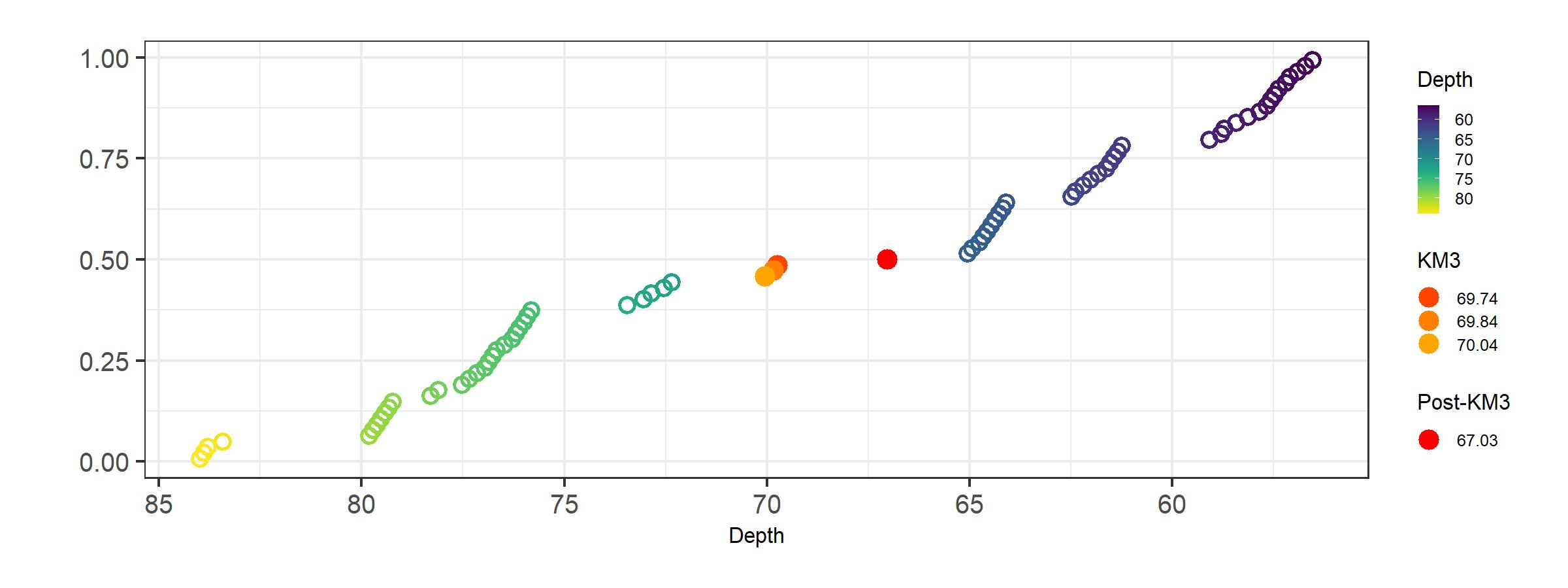}
        \caption{Depths distribution. The y axis is the equally spaced grid of the unit interval manifesting that depths form multiple clusters.}
    \end{subfigure}
    \begin{subfigure}{\textwidth}
    \centering
        \includegraphics[width=0.9\textwidth]{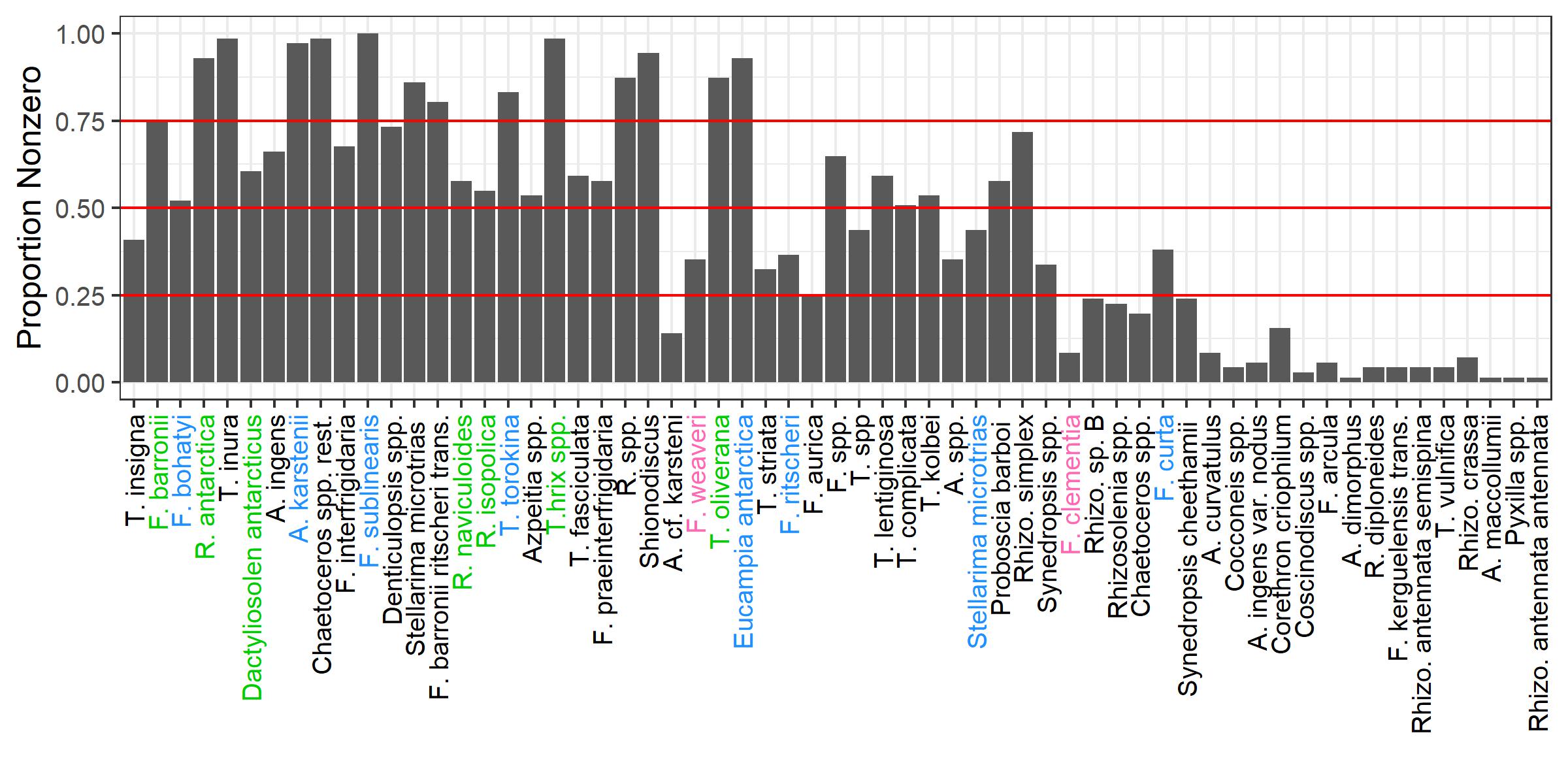}
        \caption{Distribution of the percent of nonzero observations. Some ocean temperature-related taxa are colored pink (warm-water), green (ice-tolerant open ocean), or blue (sea-ice affinity). Most observations of the last 10-15 species are zero.}
    \end{subfigure}
    \caption{Descriptive graphics of the Diatom data.}
    \label{figS:diatom distribution}
\end{figure}

In Section 6 of the main manuscript, we investigated the distribution of the compositions of the Diatom data. Figure \ref{figS:diatom distribution} (a) shows the depth distribution with an equally spaced grid of the unit interval for the y-axis. This shows that the depths where the samples were measured are not uniformly distributed but appear in clusters. Figure \ref{figS:diatom distribution}(b) shows the percent of nonzero observations for each of the diatom species. The bar plot suggests that the percent of nonzero observations is not proportional to the variance, and the most observations of the last 10-15 species are zero.

\subsection{Scores and Loading Vectors}

Figure \ref{figS:diatom score}(a)-(e) are the score plot matrices of the five methods. The same color scheme as in Figure 8 of the main manuscript was used and the outlier is marked by a red circle as before. Note that the diagonal panels are different from the density plots that are usual for scatter plot matrices. Each of the diagonal panels has the x-axis of the score and the y-axis of the depth. Figure \ref{figS:diatom loading bar} provides the corresponding loading vectors.

The detailed discussion on the scores and the loading vectors of PSA-O given in Section \ref{sec:real data} is compared to those of PSA-S. Recall that the first, second, and fourth modes of variation are driven by the first phase (Depth $>$ 67), the second phase (Depth $<$ 67), and the outlier, respectively. A similar investigation into the diagonal panels of Figure \ref{figS:diatom score}(a) suggests that each of the modes of variation captures variation that is driven by either different time periods or the outlier. The first four modes of variation of PSA-S are associated with the transition of species during the first phase, the second phase, the earlier period of the first phase, and the later period of the second phase, respectively. In addition, the second mode of variation is also associated with the outlier. This observation can be attributed to that the first loading vector of PSA-S is similar to that of PSA-O, as represented by positive weight of F. barronii and R. antarctica and negative weight of T. inura and A. karstenii. The second loading vector of PSA-S is similar to a mixture of the second and the fourth loading vectors of PSA-O, as represented by positive weight of T. insigna and Denticulopsis. spp., and negative weight of T. inura and A. karstenii. However, importantly, the second mode of PSA-S is not exactly same as the mixture as A. karstenii indicates the opposite directions, negative in the second mode of PSA-S and positive in the mixture of two modes of PSA-O. The two methods are pointing out different yet important aspects of the outlier. PSA-S pulls out A. ingense which is the most particular species in the outlier, while PSA-O extracts most of the species that are particular in the outlier with altered weights.

The diagonal panels of Figure \ref{figS:diatom score}(c) suggests that the first, second, and fourth modes of variation of PCA are associated with the transition of diatom species through the entire period, the second phase, and the later part of the first phase, respectively, while the third mode is associated with the outlier. The second and the third rows of Figure \ref{figS:diatom loading bar} indicate that the first mode of PCA is a mixture of the negative of the first mode and the positive of the second mode of PSA-O, as represented by postivie weight of T. insigna and A. karstenii and negative weight of F. barronii and R. antarctica. The second mode of PCA is a mixture of the second and the third mode of PSA-O as represented by extreme weights of T. instigna and F. bohatyi. The third mode of PCA and the fourth mode of PSA-O share some species but also have some distinct species, indicating that they see different aspects of the outlier.

The first two scores of the power transform PCA show similar trends to those of PCA except that the outlier begins to stick out in the second mode of variation. Accordingly, A. ingens is introduced in the second loading vector of the power transform PCA while the first two loading vectors of the power transform PCA are mostly the same as those of PCA. On contrary, the third mode of the power transform PCA is more noisy and does not manifest clear association with any time period. The fourth mode is driven by the outlier.

Similarly, the first two scores and the loading vectors of the log-ratio PCA are close to those of PCA, while the outlier is more clearly pulled out in the second mode of variation as a result of higher weight of A. ingens in the loading vector.

\begin{figure}[H]
    \centering
    \begin{subfigure}[b]{\textwidth}
        \centering
        \includegraphics[width=\textwidth]{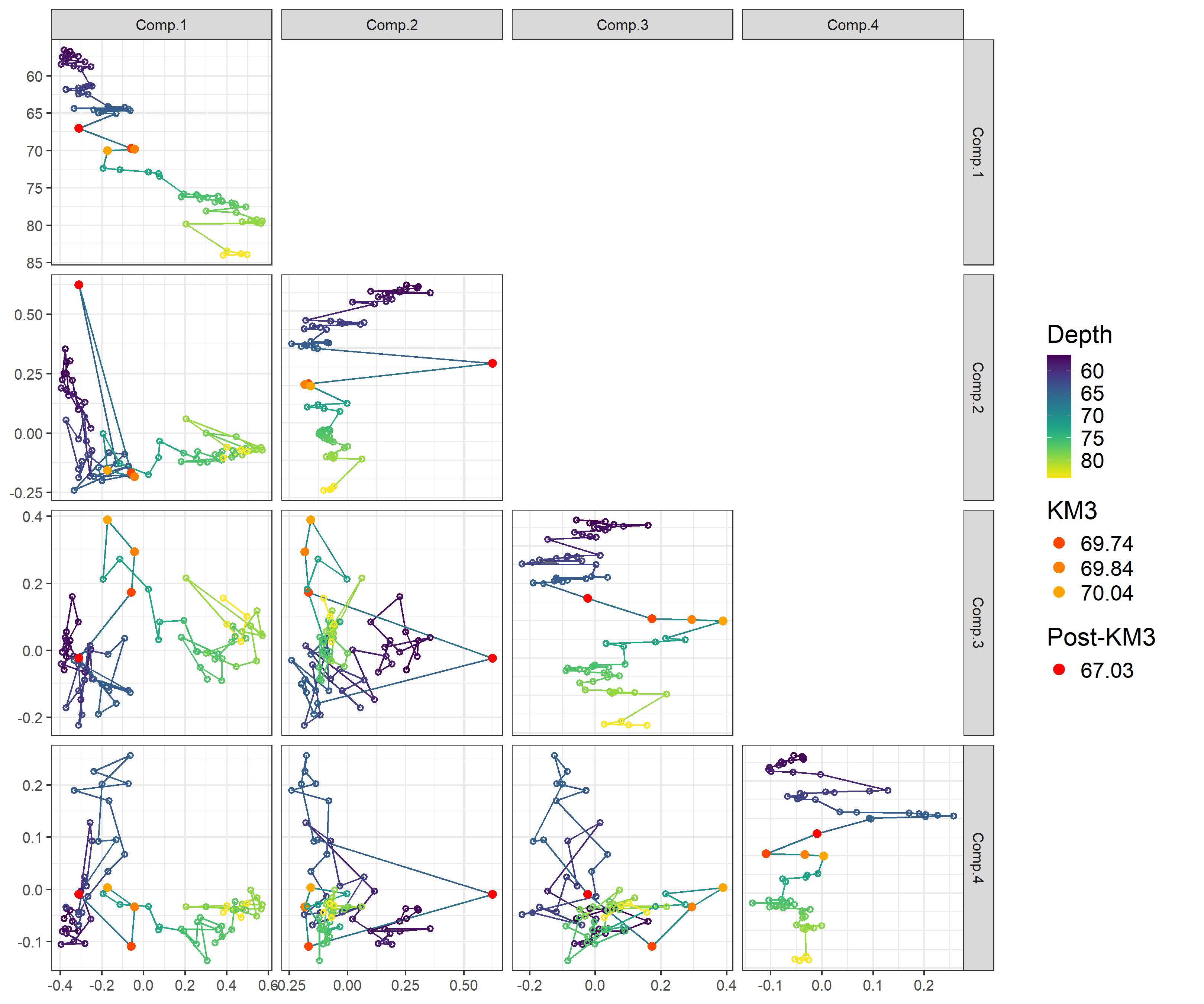}
        \caption{PSA-S scores for the Diatom data. Each mode of variation is associated with a transition of species during  different time periods. The variation in the second phase (navy) and the variation led by the outlier are compounded in the second mode of variation.}
    \end{subfigure}
    \caption{Score plot matrices of the five methods for the Diatom data.}
    \label{figS:diatom score}
\end{figure}

\begin{figure}[H]\ContinuedFloat
    \centering
    \begin{subfigure}[b]{\textwidth}
        \centering
        \includegraphics[width=\textwidth]{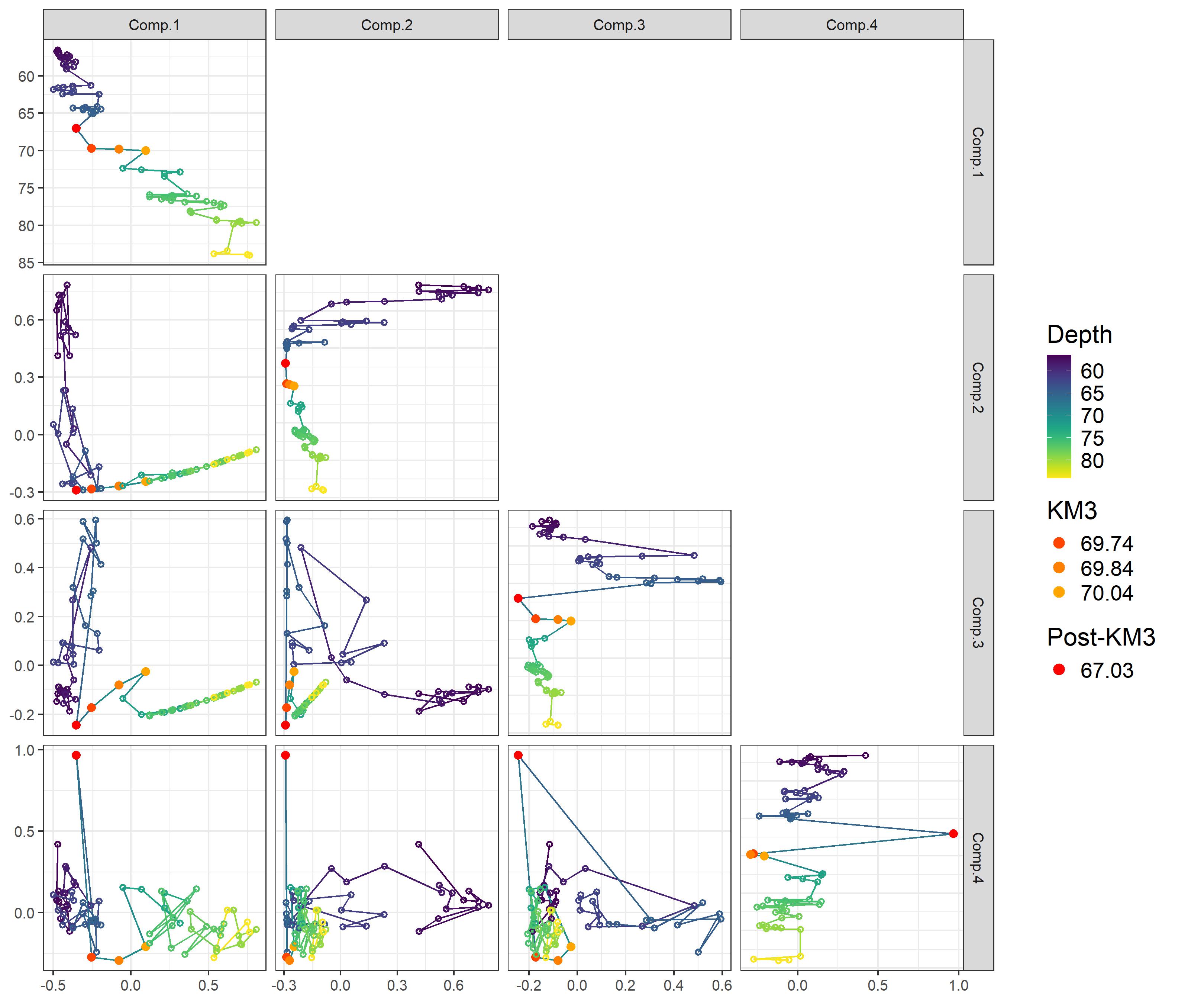}
        \caption{PSA-O scores for the Diatom data. Each mode of variation is associated with a transition of species during different time periods. The scatter plot for the first two scores highlights that the scores are orthogonal. Moreover, the two arms are parallel to the axes, facilitating an interpretation of modes of variation. The diagonal panels shows that the variation of the outlier is separated by the fourth mode of variation.}
    \end{subfigure}
    \caption{Score plot matrices of the five methods for the Diatom data.}
\end{figure}

\begin{figure}[H]\ContinuedFloat
    \centering
    \begin{subfigure}[b]{\textwidth}
        \centering  
        \includegraphics[width=\textwidth]{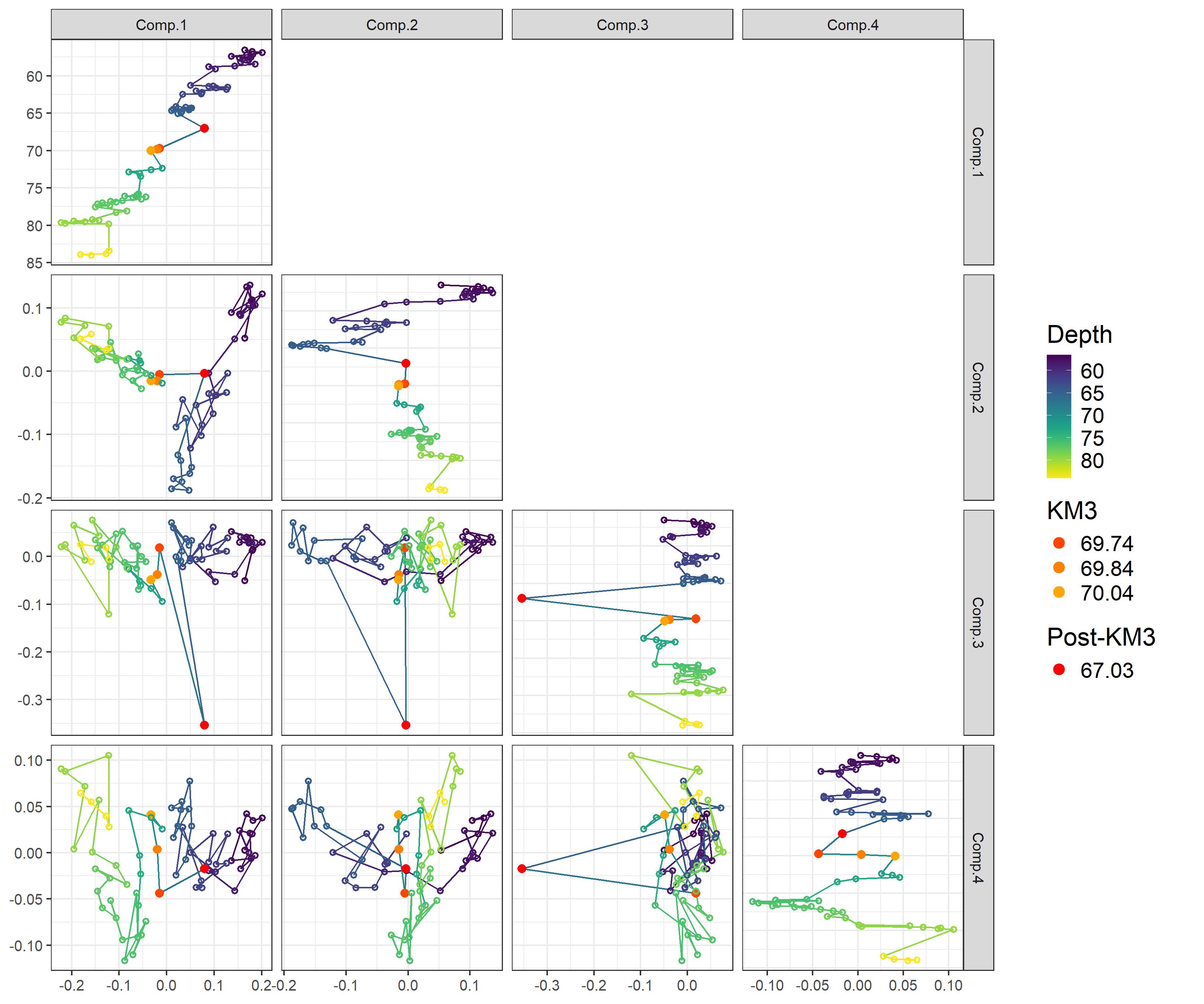}
        \caption{PCA scores for the Diatom data.}
    \end{subfigure}
    \caption{Score plot matrices of the five methods for the Diatom data.}
\end{figure}

\begin{figure}[H]\ContinuedFloat
    \centering
    \begin{subfigure}[b]{\textwidth}
        \centering  
        \includegraphics[width=\textwidth]{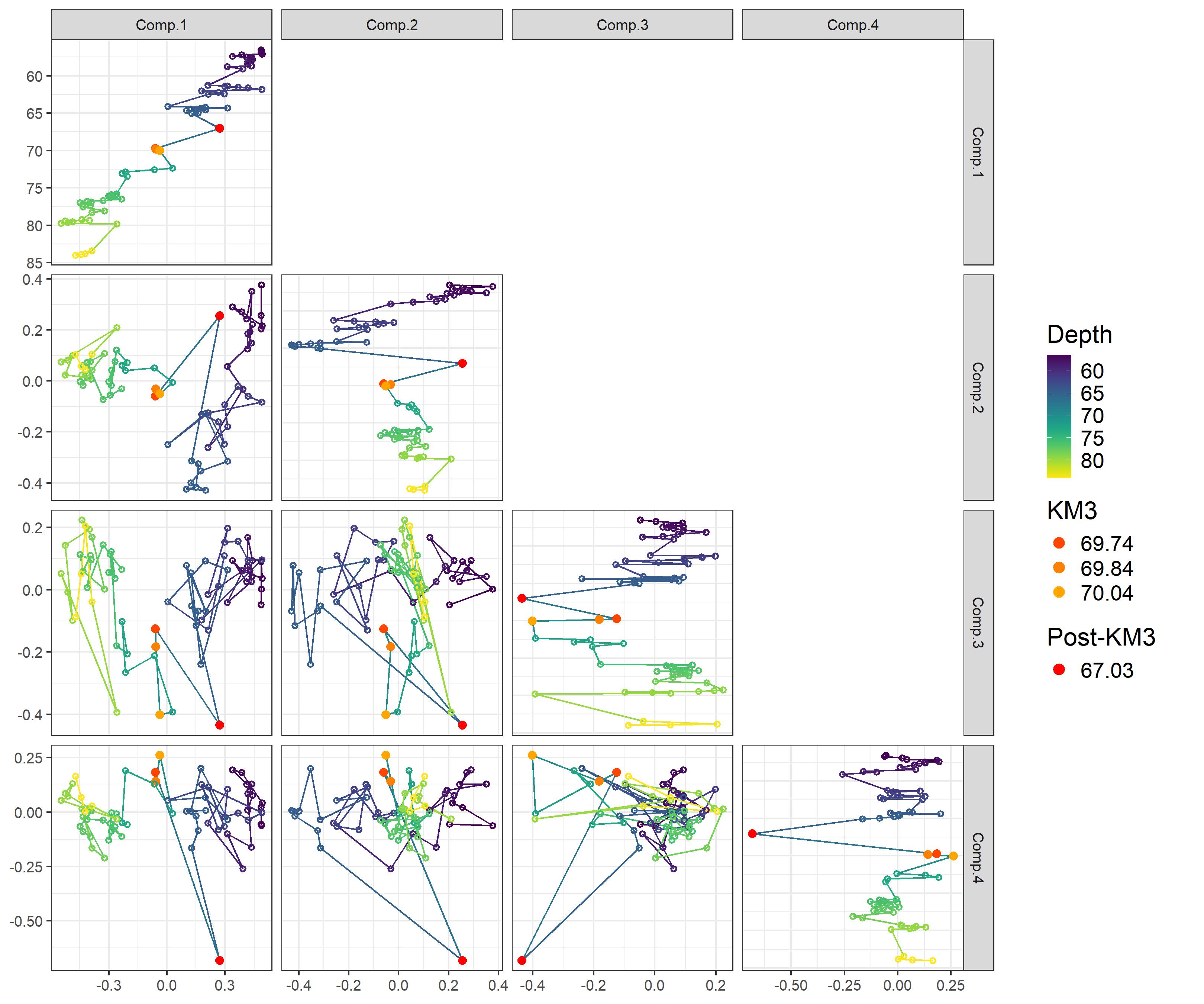}
        \caption{Power PCA scores for the Diatom data. The first two scores are similar to those of PCA, except that the second score began to capture variation of the outlier. The third score is noisy and does not indicate clear association with any time periods.}
    \end{subfigure}
    \caption{Score plot matrices of the five methods for the Diatom data.}
\end{figure}

\begin{figure}[H]\ContinuedFloat
    \centering
    \begin{subfigure}[b]{\textwidth}
        \centering  
        \includegraphics[width=\textwidth]{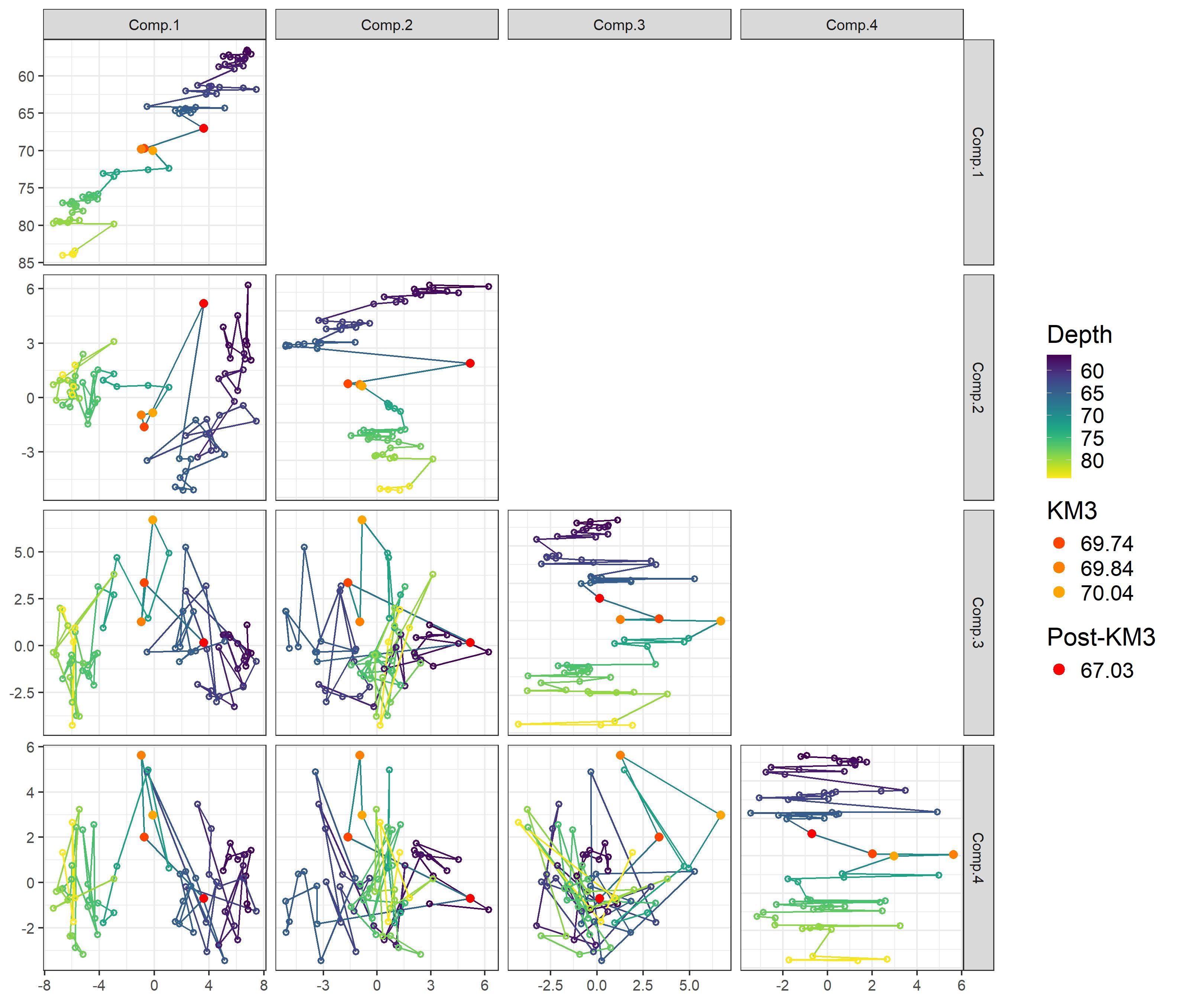}
        \caption{Log-ratio PCA scores for the Diatom data. The first two scores are similar to those of PCA, except that the second mode of variation is compounded with the outlier. The third and the fourth scores are noisy, losing clear associations with any time periods.}
    \end{subfigure}
    \caption{Score plot matrices of the five methods for the Diatom data.}
\end{figure}

\begin{figure}[H]
    \centering
    \includegraphics[width=\textwidth]{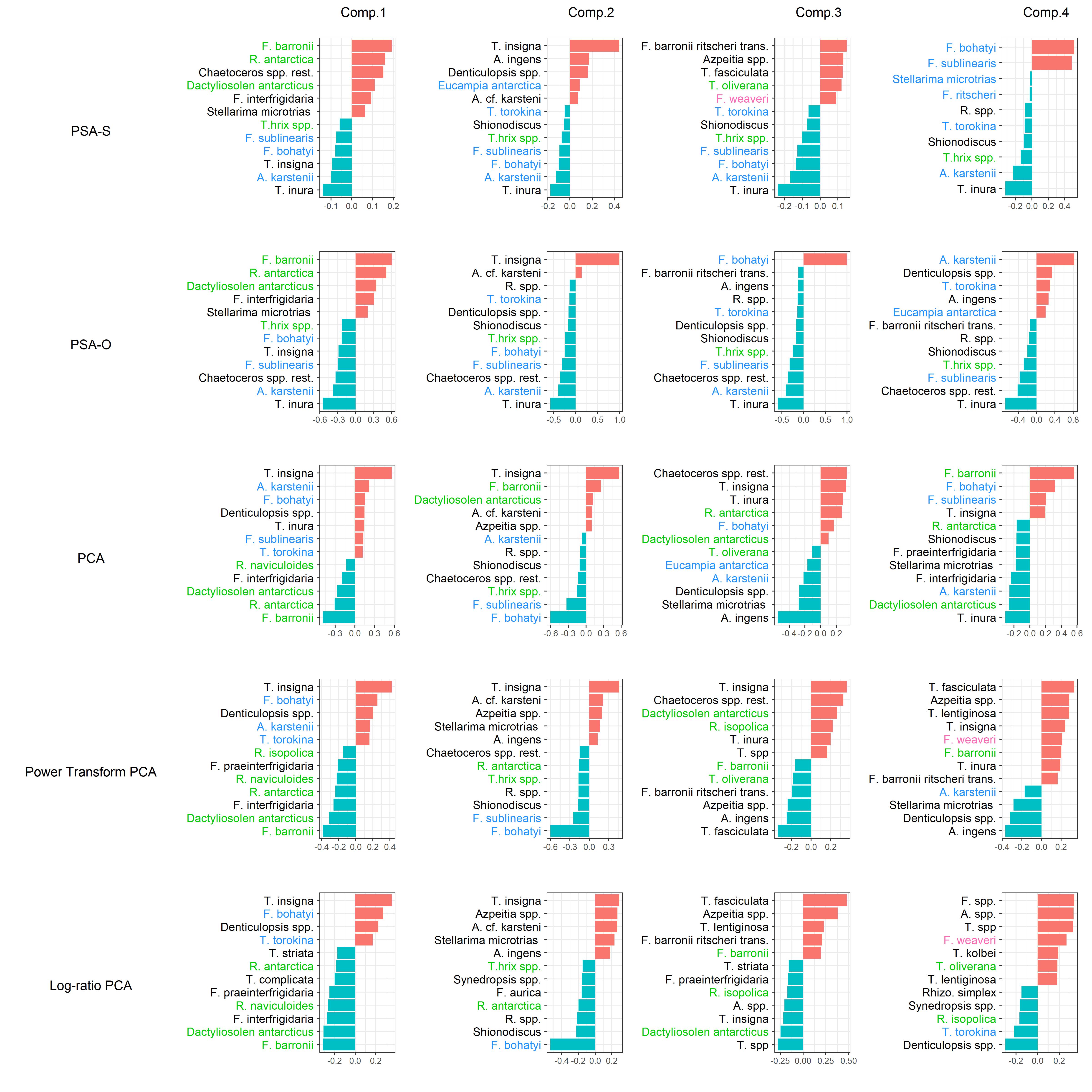}
    \caption{Bar plot representations of loading vectors of the five methods for the Diatom data. The modes of variation of different methods that are associated with the same time period are mixtures of each others, sharing the same diatom species.}
    \label{figS:diatom loading bar}
\end{figure}

\subsection{Parallel Coordinate Plot Representation of Loading Vectors}

An alternative way of representing loading vector is using parallel coordinate plots. Figure \ref{figS:diatom loading parallel} shows parallel coordinate plots for PSA-S and PSA-O loading vectors. An advantage of using parallel coordinate plots over using bar plots is that the same species appears at the same x-axis so one can easily compare the loading vectors across different methods. For example, by comparing Figure \ref{figS:diatom loading parallel}(a) and \ref{figS:diatom loading parallel}(b) one can quickly notice that the first loading vectors of PSA-S and PSA-O are similar and the second loading vector of PSA-S is a mixture of the second and the fourth loading vectors of PSA-O. On contrary, an advantage of using bar plots is that one can easily recognize the important players in the loading vectors, particularly when there are more than a handful of variables.

\begin{figure}[H]
    \centering
    \begin{subfigure}[b]{\textwidth}
        \centering
        \includegraphics[width=\textwidth]{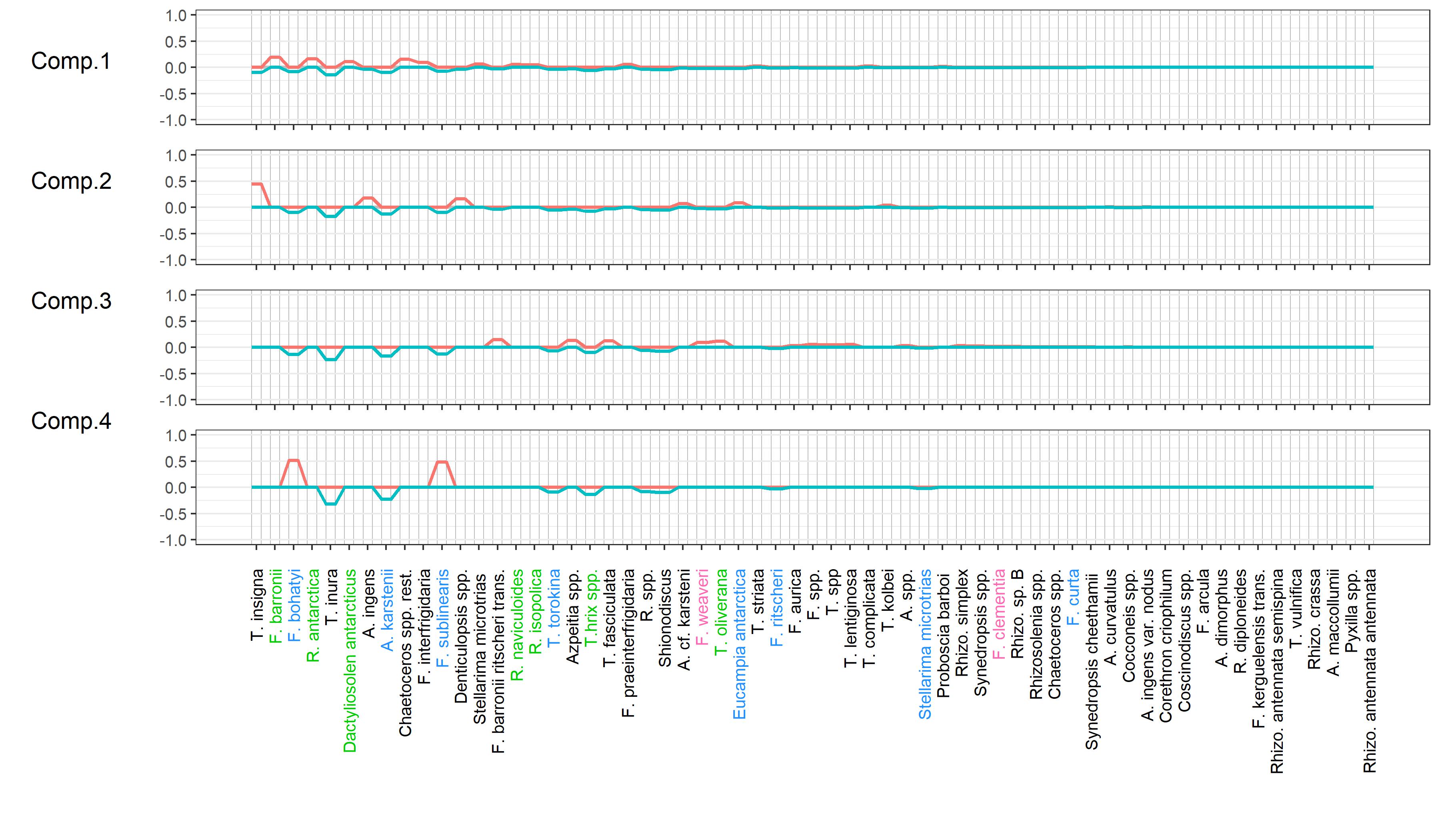}
        \caption{Loading vectors of PSA-S for the Diatom data.}
    \end{subfigure}
    \caption{Parallel coordinate plot representation of loading vectors of PSA for the Diatom data.}
    \label{figS:diatom loading parallel}
\end{figure}

\begin{figure}[H]\ContinuedFloat
    \centering
    \begin{subfigure}[b]{\textwidth}
        \centering
        \includegraphics[width=\textwidth]{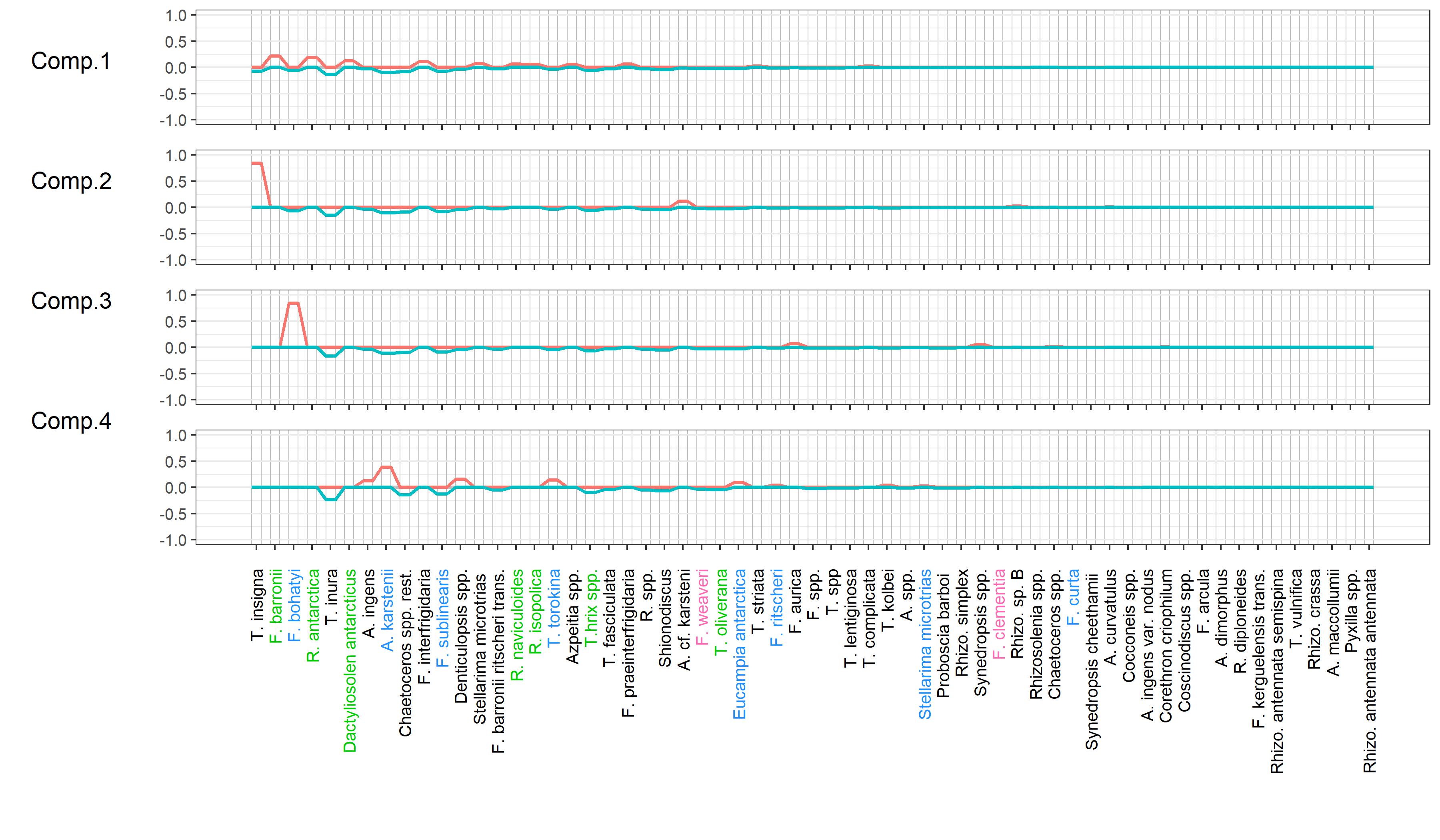}
        \caption{PSA-O loading vectors for the Diatom data.}
    \end{subfigure}
    \caption{Parallel coordinate plot representation of loading vectors of PSA for the Diatom data.}
\end{figure}

\subsection{Lower Dimensional Compositional Representations and Ternary Plots}

\begin{figure}[t!]
    \centering
    \begin{subfigure}{\textwidth}
        \centering
        \includegraphics[width = 0.6\textwidth]{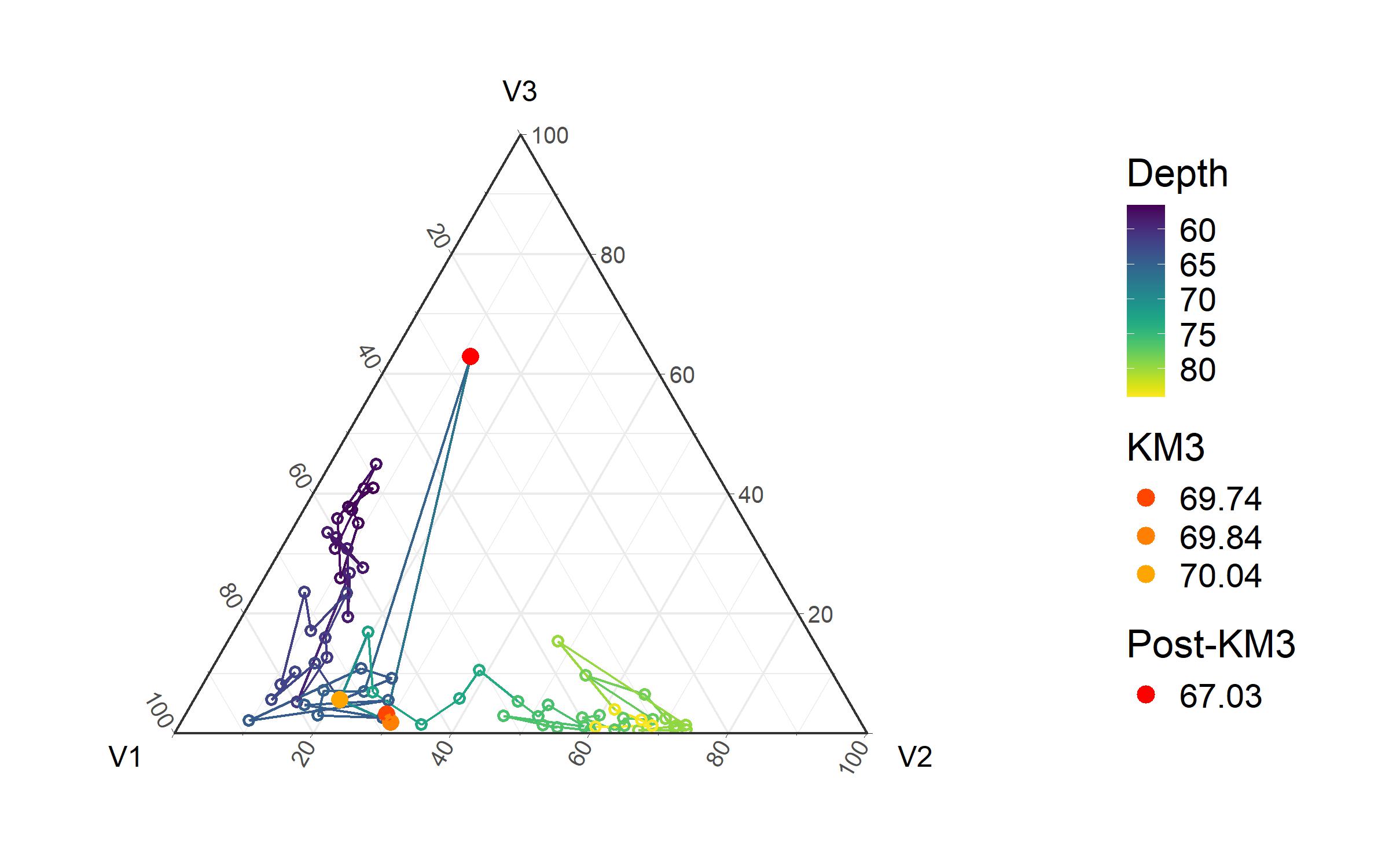}
    \end{subfigure}
    \begin{subfigure}{0.8\textwidth}
    \centering
        \includegraphics[trim = {0 1cm 0 0}, clip, width = 0.95\textwidth]{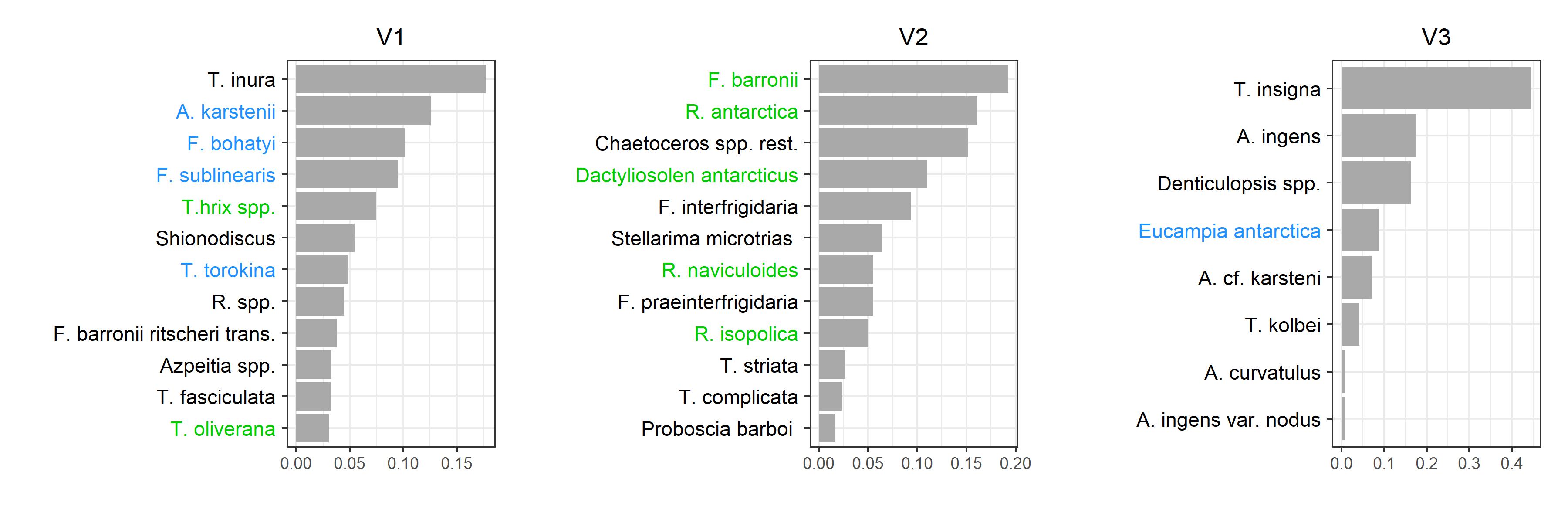}
    \end{subfigure}
    \caption{Rank 2 approximation of PSA-S for the Diatom data. The first two vertices show clear association with sea-ice affinity diatoms (blue) and open-ocean diatoms (green).}
    \label{figS:diatom ternary psas}
\end{figure}

We saw in Figure \ref{fig:diatom ternary} that the rank $2$ approximations of PSA-O can be effectively visualized through a ternary plot. The same display for PSA-S is shown in Figure \ref{figS:diatom ternary psas}. It is clear from the ternary plot that the observations in the first phase (yellow to green) have similarly low levels of $V3$, while the the observations in the second phase (green to navy) and the outlier have varied proportion of $V3$. In addition, $V3$ consists of T. insigna together with five more variables, effectively pin-pointing the variables which drive the variation in the second phase. Similarly to the vertices of PSA-O, $V1$ is more related to the sea-ice affinity diatoms and $V2$ is more related to open-ocean diatoms.

\end{document}